%% file: Hlavacek_ApJ_SPT.tex
\documentclass[apj,numberedappendix]{emulateapj}
\usepackage{amssymb,amsmath}
\usepackage{graphicx}
\usepackage{appendix}
\usepackage{natbib}
\usepackage{hyperref}
\usepackage{enumerate}
\usepackage{footnote}
\usepackage{color}

\input{defn} 

\def\aap{A\&A}                   

\def\apjs{ApJS}                  
\def\apjl{ApJ}                   

\slugcomment{}

\shorttitle{X-ray cavities in a sample of 83 SPT-selected clusters of galaxies}
\shortauthors{Hlavacek-Larrondo et al.}

\begin{document}

\title{X-ray cavities in a sample of 83 SPT-selected clusters of galaxies: \\Tracing the evolution of AGN feedback in clusters of galaxies out to $\lowercase{z}=1.2$}

\def\emailjhl{$\dagger$}
\def\UMon{1}
\def\KIPAC{2}
\def\Stanford{3}
\def\MIT{4}
\def\KICPChicago{5}
\def\FNAL{6}
\def\AAUChicago{7}
\def\CfA{8}
\def\SLAC{9}
\def\ANL{10}
\def\PhysicsUChicago{11}
\def\AIfA{18}
\def\Munich{12}
\def\ExcellenceCluster{13}
\def\Miss{14}
\def\Berkeley{15}
\def\Michigan{19}
\def\Melbourne{16}
\def\CaseWestern{17}
\def\CTIO{22}
\def\Korea{20}
\def\Hawaii{21}
\author{
J.~Hlavacek-Larrondo\altaffilmark{\UMon,\KIPAC,\Stanford,\emailjhl},
M.~McDonald\altaffilmark{\MIT},
B.~A.~Benson\altaffilmark{\KICPChicago,\FNAL,\AAUChicago},
W.~R.~Forman\altaffilmark{\CfA},
S.~W.~Allen\altaffilmark{\KIPAC,\Stanford,\SLAC},
L.~E.~Bleem\altaffilmark{\KICPChicago,\ANL,\PhysicsUChicago},
M.~L.~N.~Ashby\altaffilmark{\CfA},
S.~Bocquet\altaffilmark{\Munich,\ExcellenceCluster},
M.~Brodwin\altaffilmark{\Miss},
J.~P.~Dietrich\altaffilmark{\Munich,\ExcellenceCluster},
C.~Jones\altaffilmark{\CfA},
J.~Liu\altaffilmark{\Munich,\ExcellenceCluster},
C.~L.~Reichardt\altaffilmark{\Berkeley,\Melbourne},
B.~R.~Saliwanchik\altaffilmark{\CaseWestern}, 
A.~Saro\altaffilmark{\Munich},
T.~Schrabback\altaffilmark{\AIfA},
J.~Song\altaffilmark{\Michigan,\Korea},
B.~Stalder\altaffilmark{\CfA,\Hawaii},
A. Vikhlinin\altaffilmark{\CfA},
and
A.~Zenteno\altaffilmark{\Munich,\CTIO}
\vspace{0.1in}}
\email{Email: juliehl@astro.umontreal.ca}
\affil{$^{\UMon}${D\'{e}partement de Physique, Universit\'{e} de Montr\'{e}al, C.P. 6128, Succ. Centre-Ville, Montreal, Quebec H3C 3J7, Canada} \\
$^{\KIPAC}${Kavli Institute for Particle Astrophysics and Cosmology, Stanford University, 452 Lomita Mall, Stanford, CA 94305}\\
$^{\Stanford}${Department of Physics, Stanford University, 382 Via Pueblo Mall, Stanford, CA 94305}\\
$^{\MIT}${Kavli Institute for Astrophysics and Space Research, Massachusetts Institute of Technology, \\77 Massachusetts Avenue, Cambridge, MA 02139}\\
$^{\KICPChicago}${Kavli Institute for Cosmological Physics, University of Chicago, 5640 South Ellis Avenue, Chicago, IL 60637}\\
$^{\FNAL}${Fermi National Accelerator Laboratory, Batavia, IL 60510-0500}\\
$^{\AAUChicago}${Department of Astronomy and Astrophysics, University of Chicago, 5640 South Ellis Avenue, Chicago, IL 60637}\\
$^{\CfA}${Harvard-Smithsonian Center for Astrophysics,60 Garden Street, Cambridge, MA 02138}\\
$^{\SLAC}${SLAC National Accelerator Laboratory, 2575 Sand Hill Road, Menlo Park, CA 94025}\\
$^{\ANL}${Argonne National Laboratory, High-Energy Physics Division, 9700 S. Cass Avenue, Argonne, IL, USA 60439}\\
$^{\PhysicsUChicago}${Department of Physics, University of Chicago, 56 Ellis Avenue, Chicago, IL 60637}\\
$^{\Munich}${Department of Physics,Ludwig-Maximilians-Universit\"{a}t,Scheinerstr.\ 1, 81679 M\"{u}nchen, Germany}\\
$^{\ExcellenceCluster}${Excellence Cluster Universe,Boltzmannstr.\ 2, 85748 Garching, Germany}\\
$^{\Miss}${Department of Physics and Astronomy, University of Missouri, 5110 Rockhill Road, Kansas City, MO 64110}\\
$^{\Berkeley}${Department of Physics,University of California, Berkeley, CA 94720}\\
$^{\Melbourne}${School of Physics, University of Melbourne, Parkville, VIC 3010, Australia}\\
$^{\CaseWestern}${Physics Department, Center for Education and Research in Cosmology and Astrophysics, \\Case Western Reserve University, Cleveland, OH 44106}\\
$^{\AIfA}${Argelander-Institut f{\"u}r Astronomie, Auf dem H{\"u}gel 71, D-53121 Bonn, Germany}\\
$^{\Michigan}${Department of Physics, University of Michigan, 450 Church Street, Ann  Arbor, MI, 48109}\\
$^{\Korea}${Korea Astronomy \& Space Science Institute, Daejeon, Republic of Korea}\\
$^{\Hawaii}${Institute for Astronomy, University of Hawaii at Manoa, Honolulu, HI 96822, USA}\\
$^{\CTIO}${Cerro Tololo Inter-American Observatory, Casilla 603, La Serena, Chile}\\
\vspace{-0.1in}}

\begin{abstract}
X-ray cavities are key tracers of mechanical (or radio mode) heating arising from the active galactic nuclei (AGN) in brightest cluster galaxies. We report on a survey for X-ray cavities in 83 massive, high-redshift ($0.4<z<1.2$) clusters of galaxies selected by their Sunyaev-Zel'dovich signature in the South Pole Telescope data. Based on $Chandra$ X-ray images, we find a total of 6 clusters having symmetric pairs of surface brightness depressions consistent with the picture of radio jets inflating X-ray cavities in the intracluster medium. The majority of these detections are of relatively low significance and require deeper follow-up data in order to be confirmed.  Further, this search will miss small ($<$10\,kpc) X-ray cavities which are unresolved by $Chandra$ at high ($z\gtrsim0.5$) redshift. Despite these limitations, our results suggest that the power generated by AGN feedback in brightest cluster galaxies has remained unchanged for over half of the age of the Universe ($>7$ Gyrs at $z\sim0.8$). On average, the detected X-ray cavities have powers of $0.8-5\times10^{45}\ergps$, enthalpies of $3-6\times10^{59}\erg$, and radii of $\sim17$ kpc. Integrating over 7 Gyrs, we find that the supermassive black holes in the brightest cluster galaxies may have accreted $10^8$ to several $10^9M_{\rm \odot}$ of material to power these outflows. This level of accretion indicates that significant supermassive black hole growth may occur not only at early times, in the quasar era, but at late times as well. We also find that X-ray cavities at high redshift may inject an excess heat of $0.1-1.0$ keV per particle into the hot intracluster medium above and beyond the energy needed to offset cooling. This value is similar to the energy needed to preheat clusters, break self-similarity, and explain the excess entropy in hot atmospheres. 
\vspace{0.1in}
\end{abstract}


\keywords{Galaxies: clusters: general - X-rays: galaxies: clusters - galaxies: jets - black hole physics \vspace{-0.2in}}

\section{Introduction}
\setcounter{footnote}{0}

The observed correlations between supermassive black hole mass and galaxy bulge properties such as stellar velocity dispersion \citep[e.g.,][]{Fer2000539,Geb2000539}, spheroid mass \citep[e.g.,][]{Kor199533}, and spheroid luminosity \citep[e.g.,][]{Mag1998115}, strongly suggest that supermassive black holes and their host galaxies grew in concert with one another \citep[see][]{Ale201256}. The majority of the growth of these black holes occurs at high redshifts ($z\gtrsim1$) when they are accreting at rates near the Eddington limit, leading to powerful radiative or ``quasar-mode'' feedback \citep[e.g.,][]{Fab2012}. It is thought that this mode of feedback from active galactic nuclei (AGN) in the early universe has led to the aforementioned correlations \citep[e.g.,][]{DiM2005433,Cro2006365,Spr2005435}.

In direct contrast, local supermassive black holes are generally accreting at rates well below 1\% of the Eddington limit. A large fraction of these black holes appear to be driving powerful jetted outflows that often extend beyond the galactic hosts \citep[e.g.,][]{Kor2006372}. The low accretion rates of these black holes imply that they are most likely powered by advection-dominated accretion flows \citep[ADAFs;][and references therein]{Nar1994428}. In contrast to radiation-dominated, ``quasar-mode'' feedback, these systems are providing mechanically-dominated ``radio-mode'' feedback \citep[e.g.,][]{McN201214}. It remains unclear how and why supermassive black holes switch from one mode to the other over cosmic time.

\citet[][hereafter HL12 and HL13 respectively]{Hla2012421,Hla2013431} recently attempted to answer this question by targeting high-redshift ($z>0.3$) brightest cluster galaxies (BCGs). The AGN in these massive galaxies provide some of the most compelling cases for radio mode feedback. Jetted outflows are often seen to emerge from the AGN in BCGs and, as they propagate through the hot X-ray emitting intracluster medium (ICM), they push aside the ICM creating regions of depleted X-ray
emission known as X-ray cavities (or bubbles). These X-ray cavities provide a unique opportunity to directly measure the work done by ``radio-mode'' AGN feedback on the surrounding medium \citep[e.g.,][]{Bir2004607,Bir2008686,Dun2005364,Dun2006373,Raf2006652,Dun2008385,Nul2007,Dun2010404,Cav2010720,Don2010712,OSu2011735}. In particular, HL12 and HL13 showed that the AGN in BCGs were becoming increasingly more X-ray luminous with redshift while the mechanical properties of their outflows - as characterized by the properties of their X-ray cavities - remain unchanged. In other words, HL12 and HL13 were directly witnessing the transition between quasar mode feedback to radio mode feedback \citep[see also][]{Ued2013778,Ma2011740,Ma2013763}. 

HL12 and HL13 based their study on the MAssive Cluster Survey (MACS), a survey consisting of highly X-ray luminous clusters at $0.3<z<0.7$ \citep{Ebe2001553,Ebe2007661,Ebe2010407}, but the MACS sample contains only 4 clusters with $Chandra$ X-ray observations beyond $z=0.55$; to constrain the evolution of AGN feedback in BCGs beyond $z=0.55$, a larger sample of high-redshift clusters is required. Such measurements are possible since $Chandra$, with a point spread function of $\sim1$ arcsecond, has the potential to resolve and identify X-ray cavities out to $z\sim1.0$\footnote{We note that $Chandra$ actually holds the potential to resolve X-ray cavities of $\sim20$ kpc size out to any redshift, due to the flattening of the angular diameter distance at high redshift in $\Lambda$CDM. However, beyond $z\sim$1, observations become prohibitively expensive due to X-ray surface brightness dimming.}. Indeed X-ray cavities in massive clusters have typical radii of 15-20 kpc (see Fig. 4 in McNamara \& Nulsen 2007 and Fig. 8 in HL12)\nocite{McN200745}, corresponding to 2-2.5 arcseconds at $z=1.0$. 

\begin{figure}
\centering
\begin{minipage}[c]{0.99\linewidth}
\centering \includegraphics[width=\linewidth]{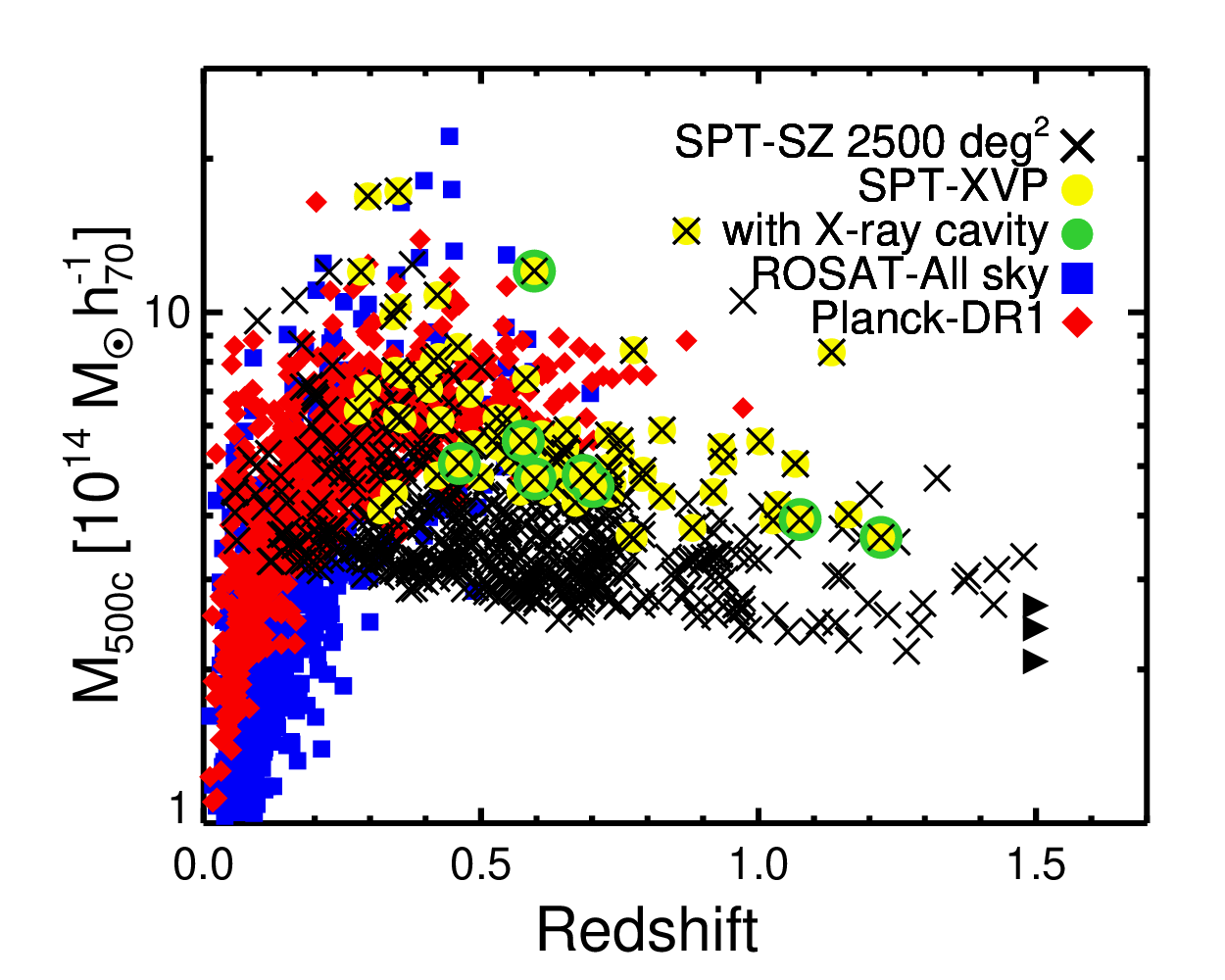}
\end{minipage}
\caption[]{Estimated cluster masses versus redshift for different cluster samples. The mass is defined as $M_{\rm 500c}$, the mass enclosed within a radius at which the average density is 500
times the critical density at the cluster redshift. We highlight in yellow the 83 SPT-SZ clusters with $Chandra$ X-ray observations that were analyzed in this study. In green, we further highlight the 8 SPT-SZ clusters with candidate X-ray cavities. The right pointing arrows symbolize the 3 high-redshift clusters at $z~\apgt~1.5$ where the Spitzer redshift model is poorly constrained.}
\label{fig1}
\end{figure} 

Recently, the number of known high-redshift galaxy clusters has increased dramatically, largely due to the success of large mm-wave surveys utilizing the Sunyaev-Zel'dovich (SZ) effect to select massive clusters at all redshifts. These SZ surveys include: the South Pole Telescope (SPT) \citep{Sta2009701,Van2010722,Rei2013763,Ble2014}, the $Planck$ satellite (Planck Collaboration 2011, Planck-29 2013)\nocite{Planck2011536,Planck201329} and the Atacama Cosmology Telescope (ACT) \citep[][]{Mar2011737,Sti2013772,Hes20137}. 
Since the surface brightness of the SZ effect is redshift independent, SZ-surveys have the potential to select nearly mass-limited cluster samples out to the earliest epochs of cluster formation. This is in contrast to previous surveys, such as those based on X-ray selection methods \citep[e.g.,][]{Gio199494,Bur2007172} which have strong redshift dependent selection functions from cosmological dimming (Fig.\ \ref{fig1}). The 2500 deg$^2$ SPT-SZ cluster survey \citep{Ble2014} in particular contains 83 massive clusters at $0.3<z<1.2$ that have now been observed with $Chandra$ \citep[e.g.,][hereafter M13]{McD2013774}. By examining the $Chandra$ X-ray images of these 83 SPT-SZ clusters, we aim to extend the sample of known clusters of galaxies with X-ray cavities out to $z\sim1$. This will allow us to determine if AGN feedback is indeed operating differently at high redshift as suggested by HL12 and HL13. 

In this paper we focus on the X-ray cavity properties - the AGN radiative properties will be explored in a future paper. In Section 2, we present the $Chandra$ X-ray data, then in Section 3 we describe the method for identifying X-ray cavities. In Section 4, we explain how we calculate cooling luminosities. Finally, in Section 5 we present the results, and discuss them in Section 6. We state the conclusions in Section 7. Throughout this paper, we adopt $H_\mathrm{0}=70\kmpspMpc$ with $\Omm=0.3$ and $\OmL=0.7$. All errors are $1\sigma$ unless otherwise noted. 

\begin{table*}
\caption{SPT-SZ clusters with candidate X-ray cavities. The first division contains the 6 systems with visually convincing cavities, while the second contains the 2 systems with marginally visually convincing cavities, see paragraph 4 in Section 3.1 for details. 1) Name; 2) Redshift; 3) Bolometric cooling luminosity within which $t_{\rm cool}$ is equal to $7.7~{\rm Gyrs}$; 4) Same but defined within $r=50$ kpc; 5) Same but defined within which $t_{\rm cool}$ is equal to the look-back time since $z=2$; 6) Cavity number (N) with the depth of the cavity compared to the surrounding X-ray emission: C for ``clear" and P for ``potential", see Section 3.1; 7) Cavity radius along the jet (errors are $\pm$20$\%$); 8) Cavity radius perpendicular to the jet (errors are $\pm$20$\%$); 9) Distance from the central AGN to the center of the cavity; 10) Position angle of the cavity for north to east; 11) Cavity enthalpy; 12) Cavity power; 13) Radio emission associated with AGN in the BCG.}
\centering
\resizebox{18cm}{!} {
\begin{tabular}{lccccccccccccc}
\hline
\hline
(1) & (2) & (3) & (4) & (5) & (6) & (7)&  (8) & (9) & (10) & (11) & (12) & (13)  \\
Name & $z$ &  $L_{\rm cool ~7.7 Gyrs}$ & $L_{\rm cool ~r=50~\rm{kpc}}$ & $L_{\rm cool ~z=2}$& N & $R_{\rm l}$ & $R_{\rm w}$ & $R$ & PA & $PV$ & $P_{\rm cav}$ & Radio   \\
& &   $10^{44}\ergps$ &  $10^{44}\ergps$ &  $10^{44}\ergps$ & &  kpc & kpc & kpc & $^o$ & $10^{59}$ erg & $10^{44}\ergps$ & \\
\hline
SPT-CLJ0000$-$5748 & 0.7019 & 7.9 & 5.2 & 5.0 &  $\#$1$^{P}$ & 18.6 & 10.7 & 37 & 140 & 1.9$\pm$0.8 & 9.9$\pm$5.0 & $\checkmark$ \\
& &   & &  & $\#$2$^{P}$ & 15.7 & 15.7 & 32 & 322 & 3.5$\pm$1.6 & 19.2$\pm$9.7 & \\
SPT-CLJ0033$-$6326 & 0.59712 & 0.7 & 0.7 & 0.3 &  $\#$1$^{P}$ & 33.3 & 20.0 & 62 & 100 & 3.3$\pm$1.5 & 7.7$\pm$3.9 & $\checkmark^a$ \\
& &    &  & &$\#$2$^{P}$ & 30.7 & 23.3 & 73 & 280 & 3.2$\pm$1.4 & 5.6$\pm$2.8 & \\
SPT-CLJ0509$-$5342 & 0.4607 & 1.3 & 1.5 & 0.9 &   $\#$1$^{C}$ & 16.9 & 16.9 & 32 & 42  & 1.9$\pm$0.8 & 6.8$\pm$3.4  &  $\times^b$  \\
&  &      &  & &$\#$2$^{C}$ &  24.9& 13.4 & 35 &  223 & 1.8$\pm$0.8 & 7.1$\pm$3.6 &    \\
SPT-CLJ0616$-$5227 & 0.6838 & 1.9  & 1.6 & 0.2 &  $\#$1$^{C}$ & 24.0 & 17.0 & 62 & 82 &  1.7$\pm$0.8 & 2.4$\pm$1.2 & $\checkmark$    \\
&  &    &  & & $\#$2$^{P}$ & 24.0 & 17.0 & 37 & 234 &  2.2$\pm$1.0 & 5.2$\pm$2.6 & \\
SPT-CLJ2331$-$5051 & 0.576 & 5.5  & 3.6 & 3.8 & $\#$1$^{P}$ & 13.7 & 10.5 & 27 & 59 & 1.1$\pm$0.5 & 6.4$\pm$3.2 &  $\checkmark$    \\
&  &    &  & & $\#$2$^{P}$ & 12.4 & 9.2 & 20 & 222 & 1.6$\pm$0.7 & 21.3$\pm$10.7 &   \\
SPT-CLJ2344$-$4243$^{c}$ & 0.596 & 157.0  &  117.7& 144.1 &  $\#$1$^{P}$ & 9.3 & 9.3 & 20 & 115 & 6.3$\pm$2.9 & 84$\pm$42  & $\checkmark$   \\
&  &      &  & & $\#$2$^{P}$ & 10.0 & 10.0 & 20 & 356 & 7.8$\pm$3.6 & 107$\pm$54  &  \\
\hline
SPT-CLJ0156$-$5541 & 1.2 & $<0.02$  & 1.2 & $<0.02$ &  $\#$1$^{P}$ & 20.9 & 20.9 & 40 & 139 & 2.5$\pm$1.1 & 3.8$\pm$1.9 & $\checkmark^a$ \\
&  &  &    & &$\#$2$^{P}$ & 21.2 & 39.0 & 40 & 335 & 4.6$\pm$2.1 & 9.8$\pm$4.9 &  \\
SPT-CLJ2342$-$5411 & 1.075 & 3.1 & 1.8 &  0.7 &  $\#$1$^{P}$ & 13.9 & 13.9 &  26 & 196 & 0.9$\pm$0.4 & 2.8$\pm$1.4 &   $\times$ \\
\hline
\hline
\end{tabular}}
$^a$Radio source not centered on the BCG. $^b$Dynamic range limited by the presence of a nearby, un-associated, bright radio source. $^c$Phoenix cluster \citep[e.g.,][]{McD2012Nat}.
\label{tab1}
\end{table*}

\section{X-ray observations and data reduction}

The majority of the SPT-SZ clusters with $Chandra$ X-ray data were observed through a $Chandra$ X-ray Visionary Project (XVP, PI Benson) targeting the most significant ($\xi > 6.5$) SZ detections in the first 2000 deg$^2$ of the 2500 deg$^2$ SPT-SZ survey at $z > 0.4$ \citep{Ben2013763}. The XVP exposure times were chosen to obtain $\sim$2000 X-ray counts per cluster, predicted using an SPT-significance to X-ray luminosity relation from a subset of clusters with earlier $Chandra$ observations. While a total of 80 clusters were observed through this XVP, we only consider the 74 that were observed with $Chandra$ and not the 6 that were observed with $XMM-Newton$, since $Chandra$ is the only telescope that can resolve X-ray cavities at high redshifts. In addition to these 74 clusters, we include nine SPT-selected clusters at $z > 0.3$ that were previously observed through other $Chandra$ programs. 

The final sample therefore consists of 83 massive clusters, spanning a redshift range of $0.3 \le z \le 1.2$. These clusters all have highly-significant SPT detections ($\xi > 6.5$), and we expect the SZ-selection to be nearly mass independent with no significant detection bias towards clusters with highly-peaked surface-brightness distributions \citep[][]{Mot2005623}. 

The majority of the X-ray data reduction and analysis is described in M13, to which we refer the reader for a more detailed description. Briefly, surface brightness profiles are measured in a series of 20 annuli out to 1.5$\times{R_{\rm 500}}$. These are then expressed as projected emission measure integrals of the gas density, and the latter are modeled using modified beta models \citep{Vik2006640}. With only $\sim$2000 X-ray counts per cluster, one cannot apply the standard deprojection techniques \citep[e.g.,][]{Vik2006640,Sun2009693}. Instead M13 models the underlying dark matter distributions as generalized Navarro-Frenk-White profiles \citep[GNFW;][]{Zha1996278,Wyi2001555}. Under the assumption of hydrostatic equilibrium, the authors then produce best-fit deprojected gas density and temperature profiles, along with a model for the underlying gravitational potential for each cluster. While we use these best-fit deprojected profiles throughout the paper, we note that the uncertainties derived in M13 do not reflect the underlying assumptions: typical uncertainties are significantly larger, on the order of $25\%$ in pressure and $50\%$ in the depth of the gravitational potential at the radii of cavities (see Appendix A of M13). Since the local pressure and gravitational potential are two quantities needed to compute X-ray cavity energetics, the reported cavity energetics in the SPT-SZ sample are, at best, within a factor of a few of their true value. We therefore proceed with the search of cavities in the sample while using the values derived in M13, but bear in mind throughout the paper that the uncertainties are significantly larger than reported in M13.

\begin{figure*}
\centering
\begin{minipage}[c]{0.945\linewidth}
\centering \includegraphics[width=\linewidth]{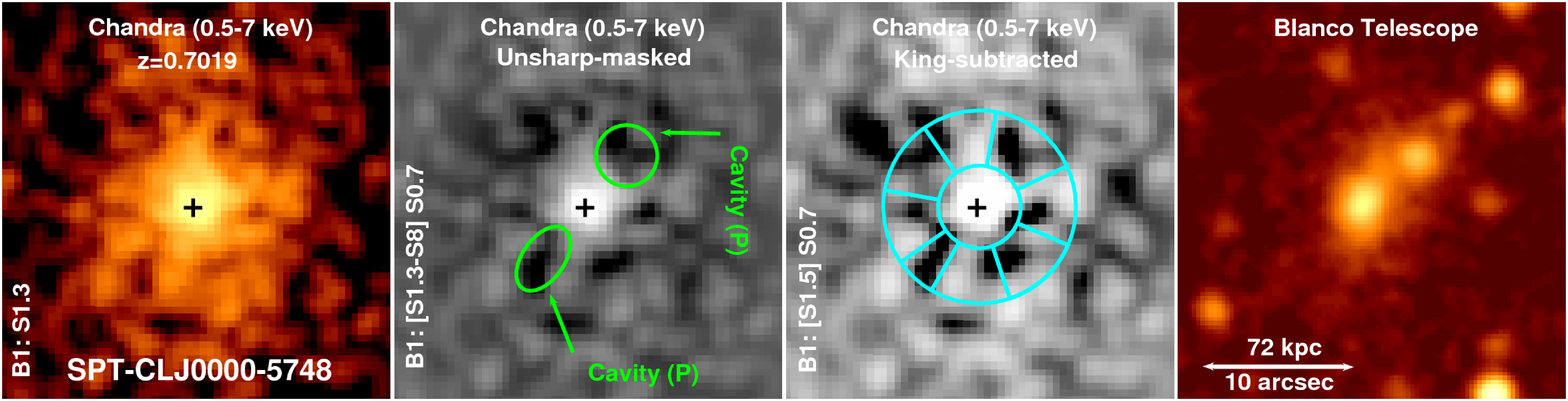}
\vspace{0.0005cm}
\end{minipage}
\begin{minipage}[c]{0.949\linewidth}
\centering \includegraphics[width=\linewidth]{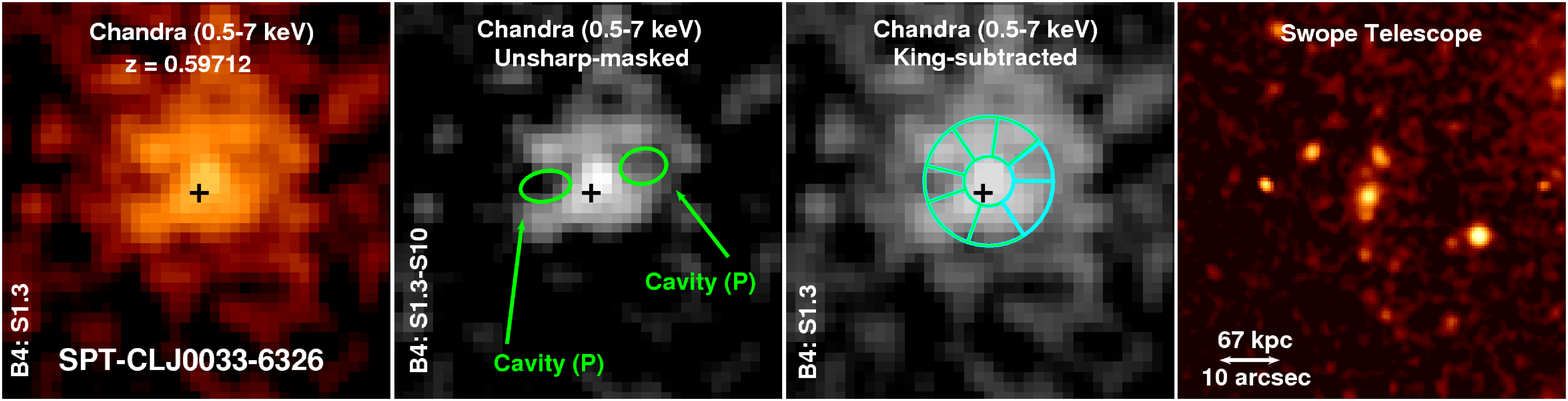}
\vspace{0.001cm}
\end{minipage}
\begin{minipage}[c]{0.949\linewidth}
\centering \includegraphics[width=\linewidth]{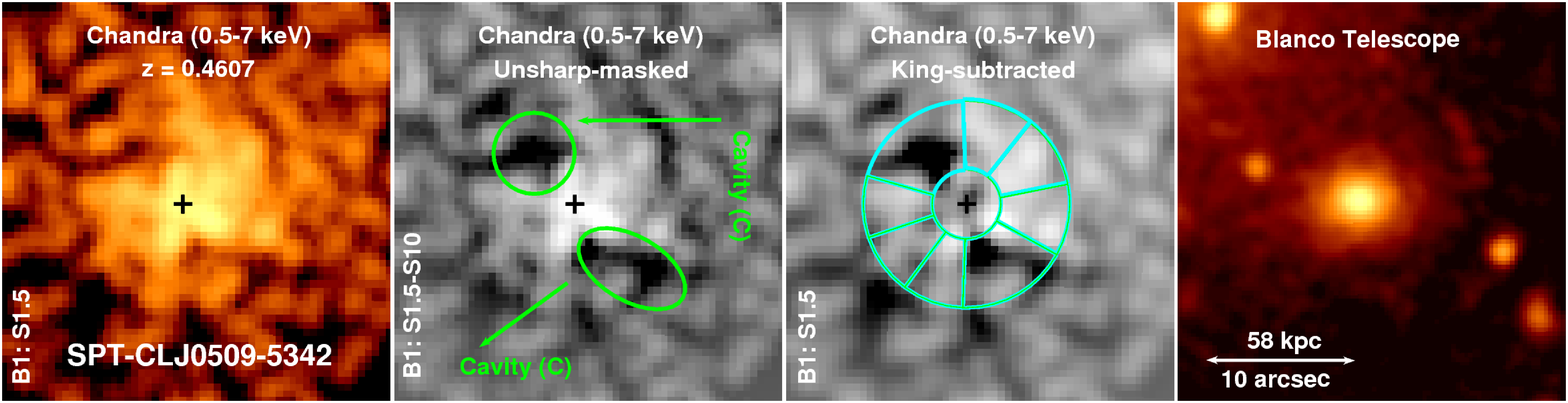}
\vspace{0.001cm}
\end{minipage}
\caption{Images of the SPT-SZ clusters with candidate X-ray cavities. For each row, we show the candidate cluster's: $0.5-7$ keV X-ray image, the $0.5-7$ keV unsharp-masked image, the $0.5-7$ keV King-subtracted image, and the optical or infrared image. The position of the central AGN, taken to be the position of the BCG, is shown with the black cross. The smoothing and binning factors are shown in the lower-left corners. We also highlight the cavities in green, as well as their depth: ``C" (``P") for clear (potential), see Section 3.1. In light blue, we illustrate the annuli used to compute the azimuthal surface brightness profiles (Fig. \ref{fig2b}). }

\end{figure*}

\setcounter{figure}{1}
\begin{figure*}
\centering
\begin{minipage}[c]{0.949\linewidth}
\centering \includegraphics[width=\linewidth]{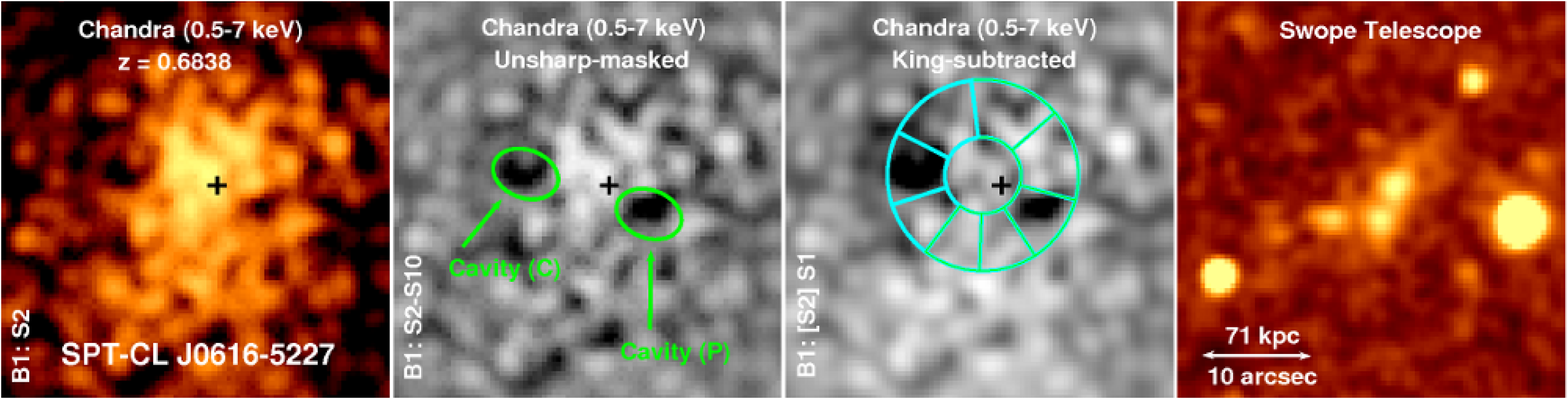}
\vspace{0.001cm}
\end{minipage}
\centering
\begin{minipage}[c]{0.949\linewidth}
\centering \includegraphics[width=\linewidth]{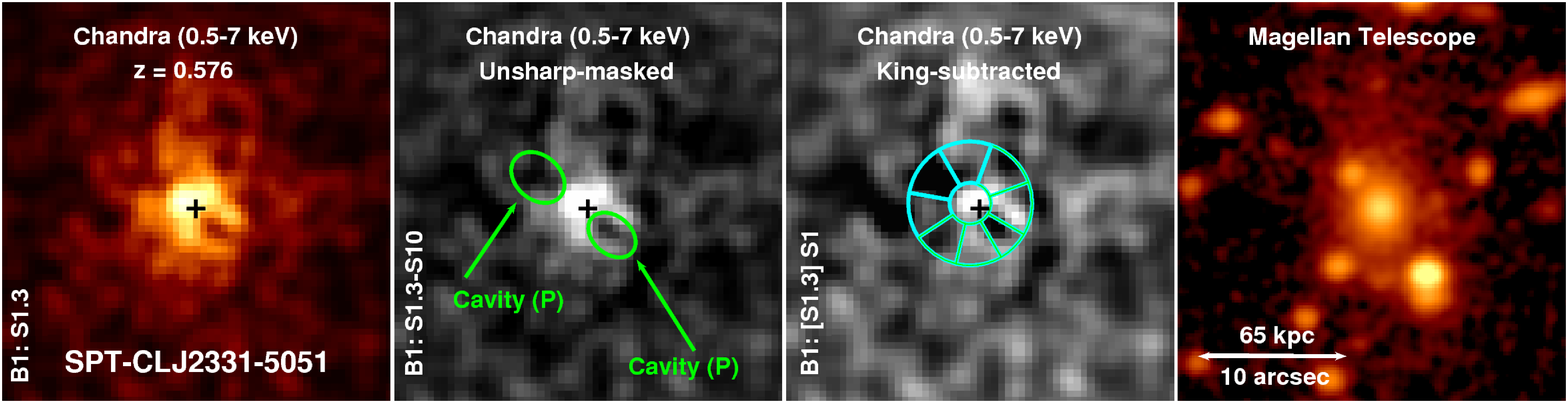}
\vspace{0.001cm}
\end{minipage}
\begin{minipage}[c]{0.949\linewidth}
\centering \includegraphics[width=\linewidth]{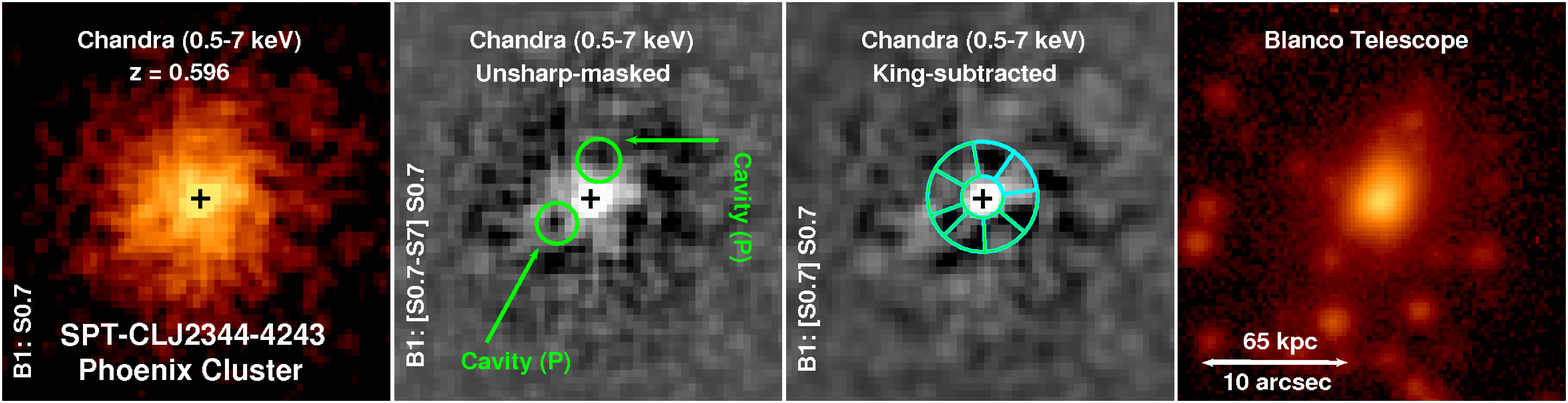}
\end{minipage}
\caption{Images of the SPT-SZ clusters with candidate X-ray cavities. See caption on page 4.}
\end{figure*}
\begin{figure*}

\setcounter{figure}{1}
\centering
\begin{minipage}[c]{0.949\linewidth}
\centering \includegraphics[width=\linewidth]{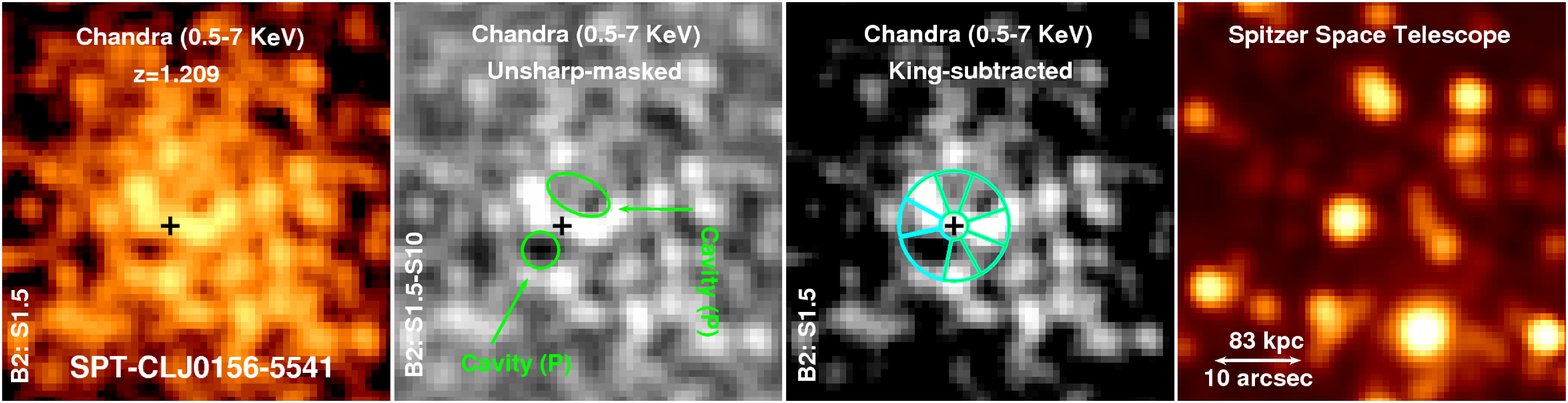}
\vspace{0.001cm}
\end{minipage}
\begin{minipage}[c]{0.949\linewidth}
\centering \includegraphics[width=\linewidth]{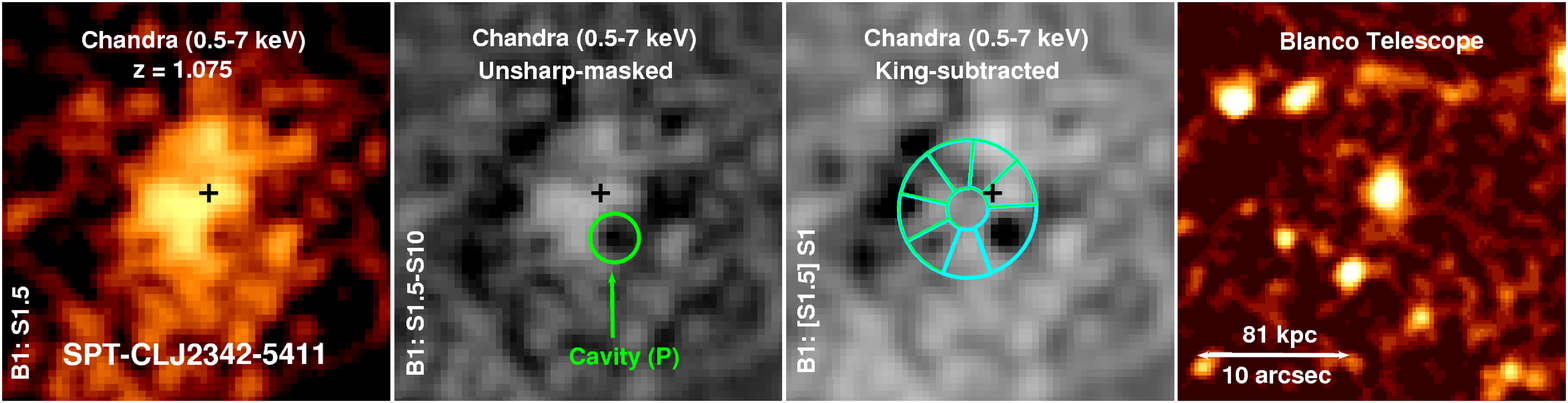}
\vspace{0.001cm}
\end{minipage}
\caption[]{Images of the SPT-SZ clusters with candidate X-ray cavities. See caption on page 4.}
\label{fig2}
\end{figure*}

\begin{figure*}
\centering
\begin{minipage}[c]{0.24\linewidth}
\includegraphics[width=\linewidth]{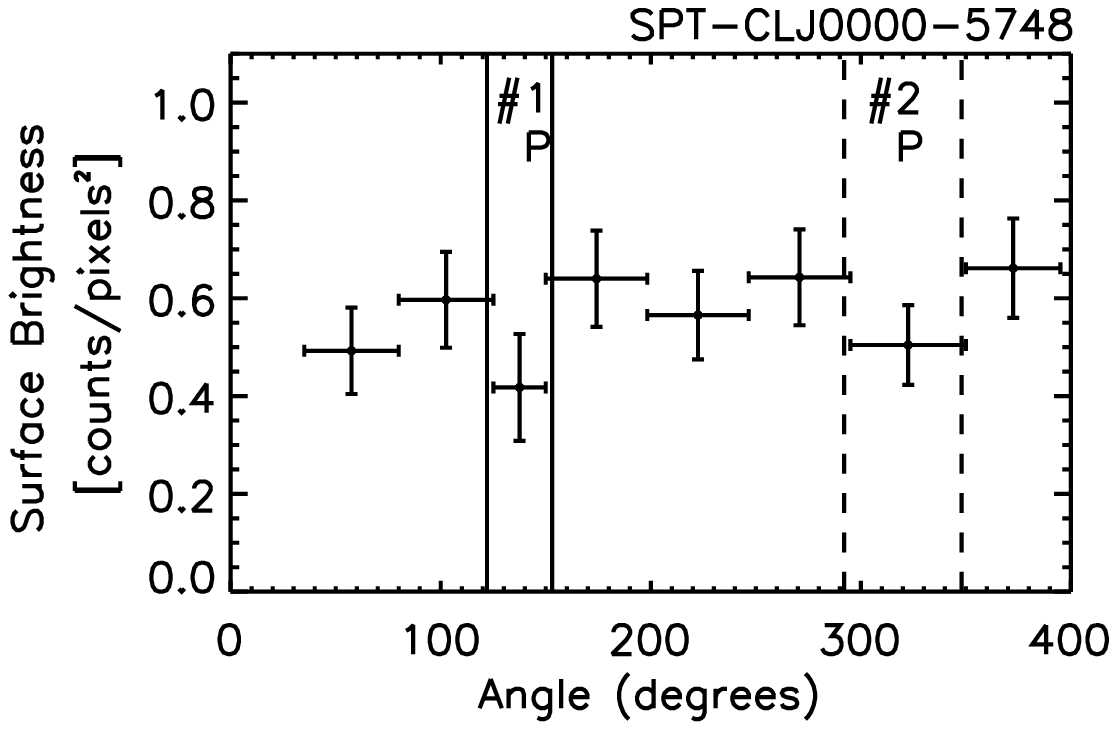}
\end{minipage}
\begin{minipage}[c]{0.24\linewidth}
\includegraphics[width=\linewidth]{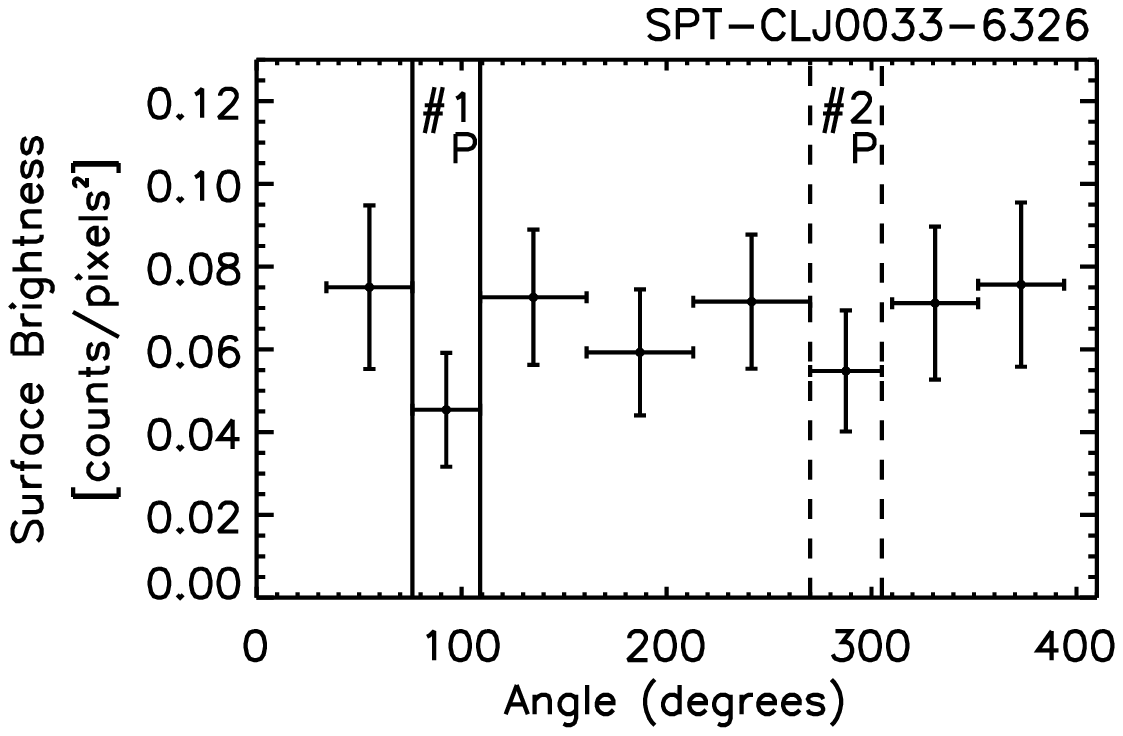}
\end{minipage}
\begin{minipage}[c]{0.24\linewidth}
\includegraphics[width=\linewidth]{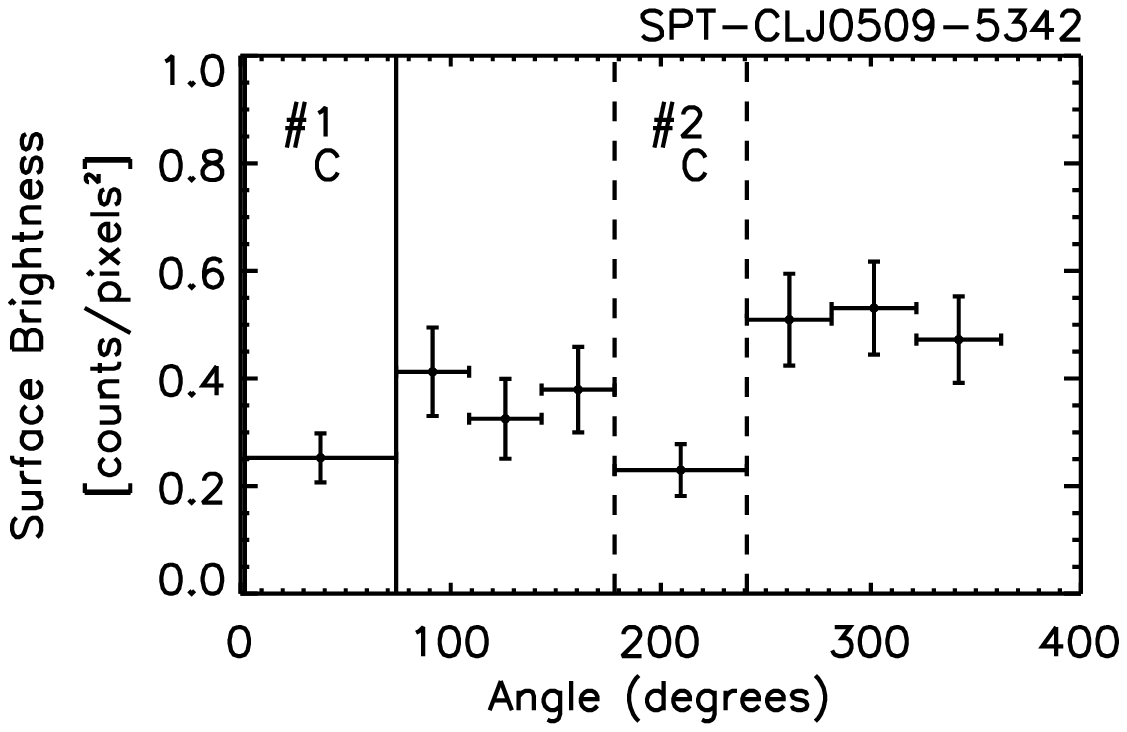}
\end{minipage}
\begin{minipage}[c]{0.24\linewidth}
\includegraphics[width=\linewidth]{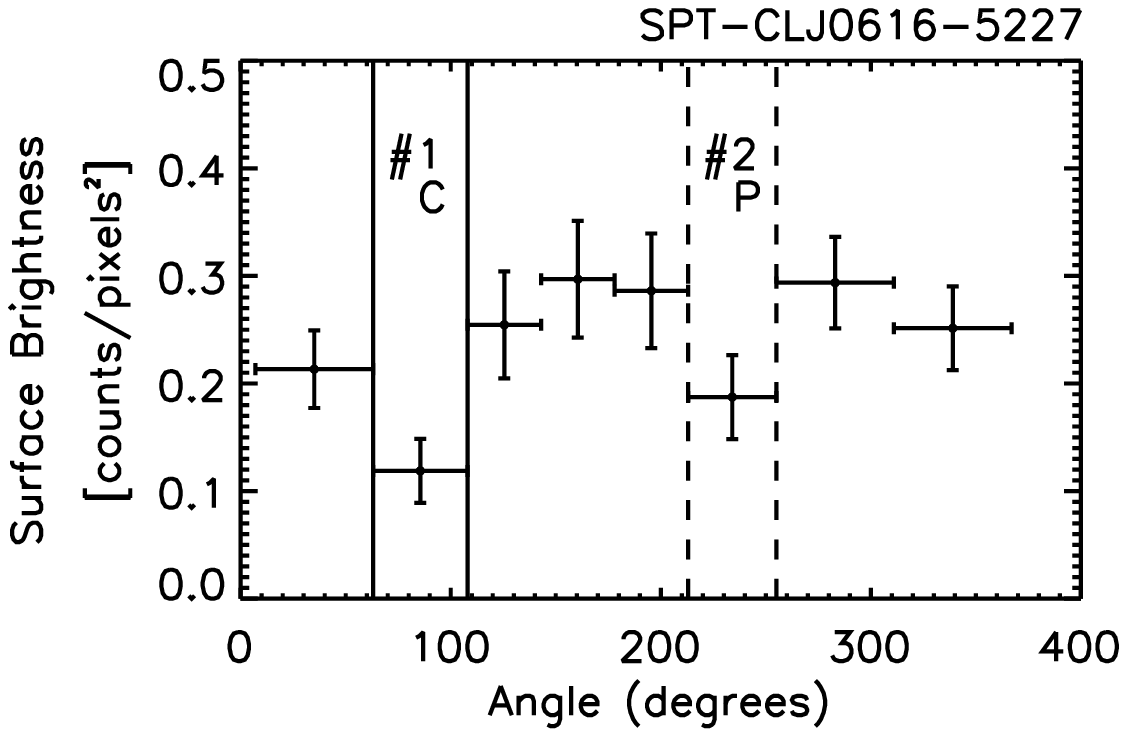}
\end{minipage}
\begin{minipage}[c]{0.24\linewidth}
\includegraphics[width=\linewidth]{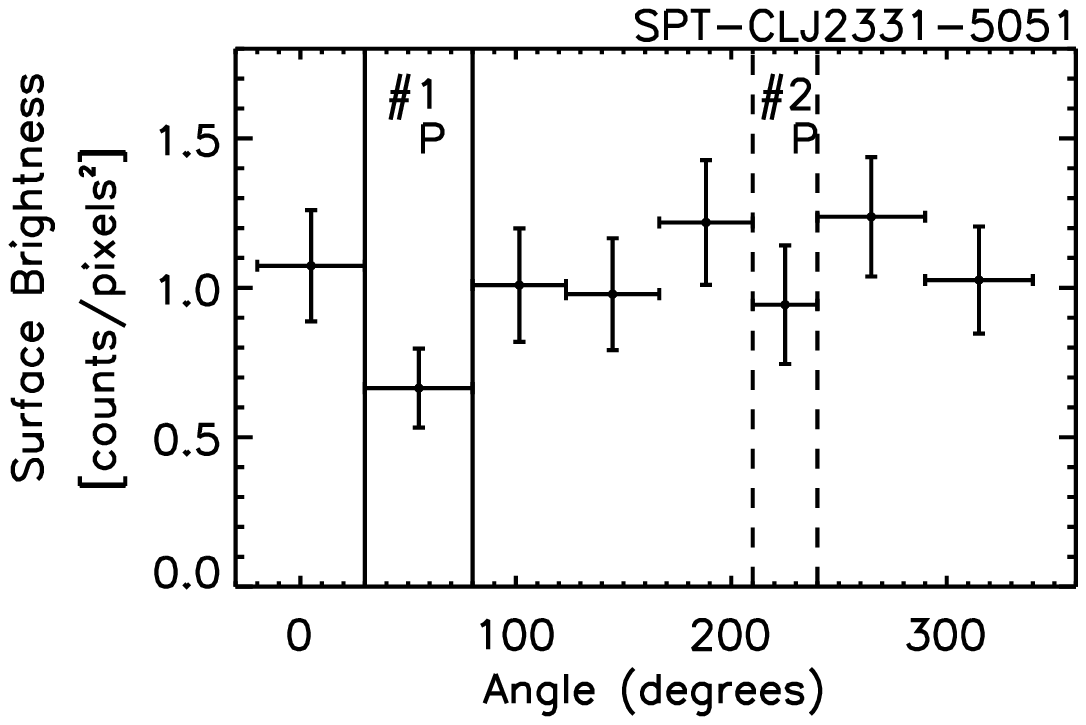}
\end{minipage}
\begin{minipage}[c]{0.24\linewidth}
\includegraphics[width=\linewidth]{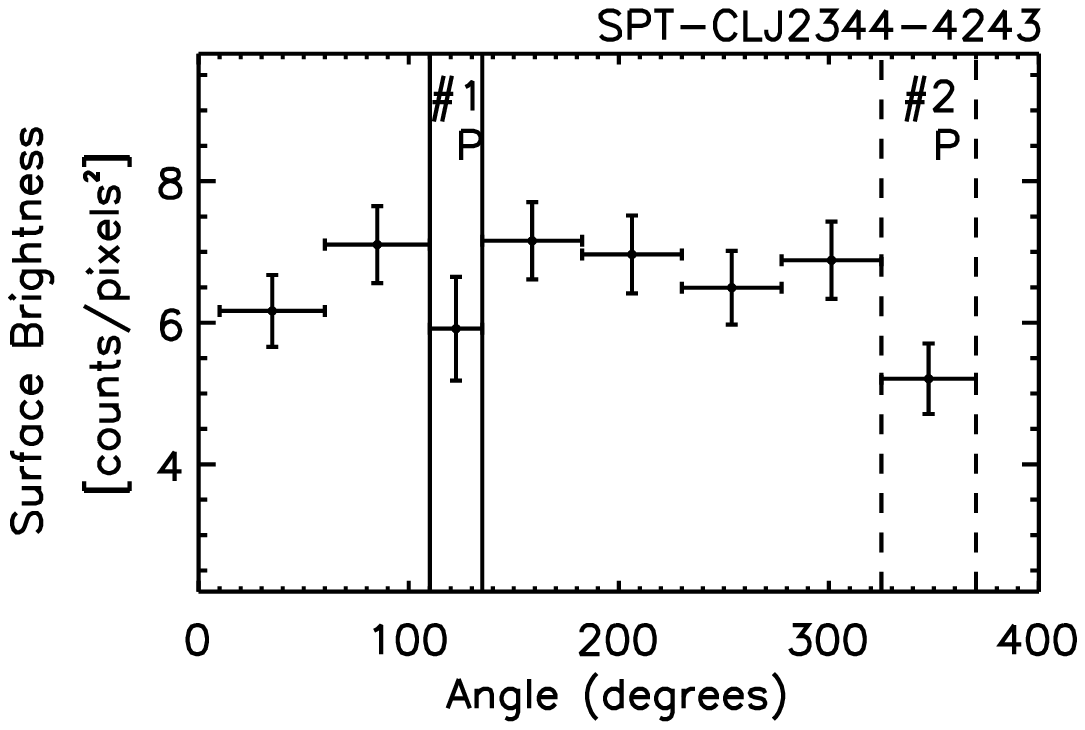}
\end{minipage}
\begin{minipage}[c]{0.24\linewidth}
\includegraphics[width=\linewidth]{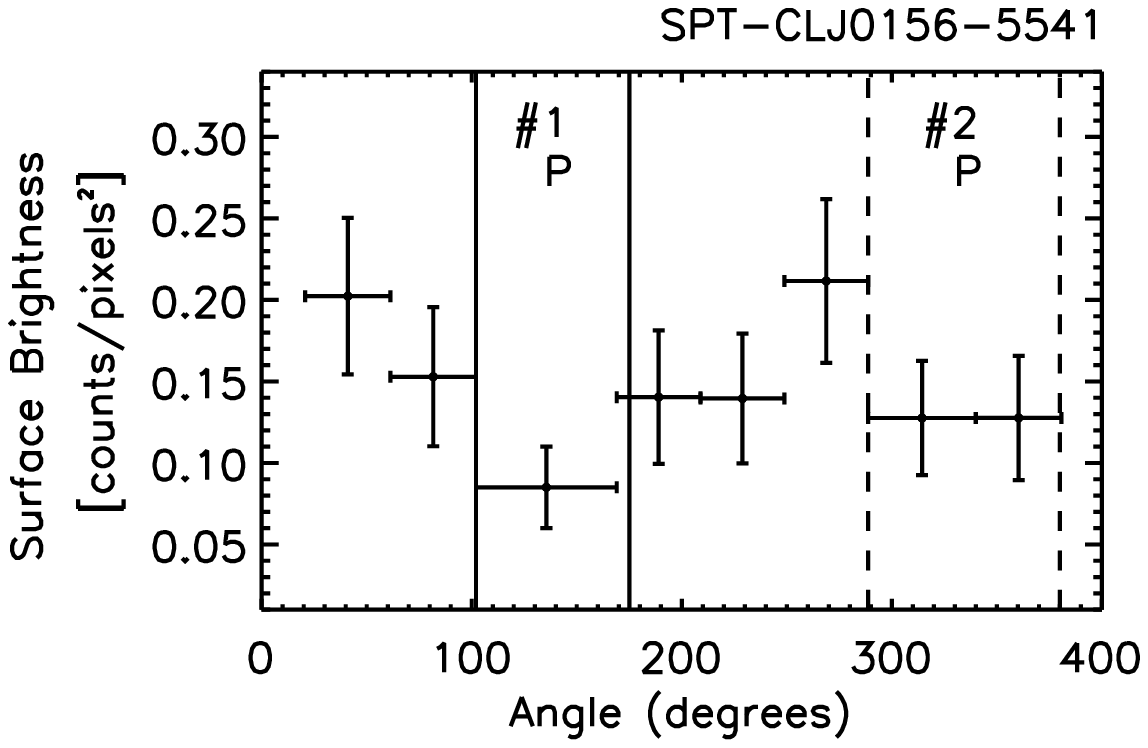}
\end{minipage}
\begin{minipage}[c]{0.24\linewidth}
\includegraphics[width=\linewidth]{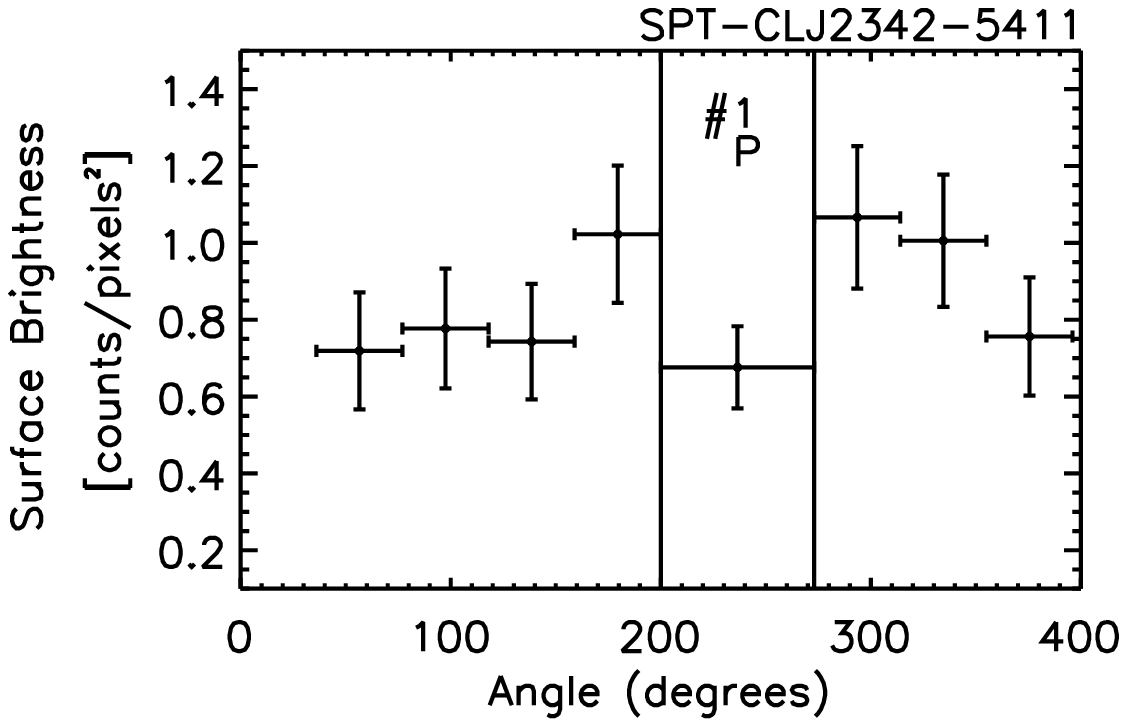}
\end{minipage}
\caption[]{We show the $0.5-7\keV$ azimuthal surface brightness profile of the annuli containing the candidate cavities (Fig. \ref{fig2}). The vertical lines illustrate the locations of each cavity. These plots show that the majority of the X-ray cavities lie $\sim1-3\sigma$ below the surrounding X-ray emission. We also indicate the depth of the X-ray cavities: ``C" (``P") for clear (potential), see Section 3.1. }
\label{fig2b}
\end{figure*}

\section{Identifying systems with X-ray cavities}

\subsection{Method}
To identify systems with X-ray cavities, we visually search the central 100 kpc of each cluster for circular or ellipsoidal surface brightness depressions in the $Chandra$ X-ray images. Past studies on local clusters have found that the majority of X-ray cavities are located within this region \citep[e.g.,][HL12]{McN200745}.

As a first indicator, we create unsharp-masked images for each cluster to enhance deviations in the original $Chandra$ X-ray image. This method consists of subtracting a strongly smoothed image from a lightly smoothed image and has been used extensively in the literature for X-ray cavity studies \citep[e.g.,][]{San2009393,Bla2009697,Mac2011743}. For the strongly smoothed image, we use gaussian smoothing scales on the order of 40 to 80 kpc to match the underlying large-scale cluster emission, and for the lightly smoothed image, we use gaussian smoothing scales on the order of 5 to 10 kpc to match the typical length scale of a cavity. Next, we build a best-fitting King model, centered on the X-ray peak, for the surface-brightness distribution of all clusters showing hints of depressions. We use the lightly smoothed $Chandra$ X-ray image to build the King model, and subtracted the resulting King model image from the original $Chandra$ X-ray image. We refer to these as the ``King-subtracted" images. The King model is defined as:
\begin{equation}
I(r) = I_{\rm 0} (1 + \left(\frac{r}{r_{\rm 0}}\right)^2)^{-\beta} + C_{\rm 0} ,
\label{eq1}
\end{equation}
where $I_{\rm 0}$ is the normalization, $r_{\rm 0}$ is the core radius, and $C_{\rm 0}$ is a constant. We allow all three of these parameters to vary while building the King model. 

We try several binning and smoothing factors for all clusters, and also test the robustness of the King-subtracted method by shifting the position of the central peak by $\pm2-3$ pixels. We examine three energy bands: $0.5-7\keV$, $0.3-2\keV$ and $0.6-3\keV$. We only consider cavities to be real if they are visually identifiable in all X-ray images including the original X-ray images (excessive binning and smoothing eventually removes any indication of a depression). 

Out of the 83 SPT-SZ clusters with $Chandra$ X-ray observations, we find that 10 clusters have surface brightness depressions in their $Chandra$ X-ray images. However, to minimize false-identification, three co-authors (Hlavacek-Larrondo, McDonald and Allen) independently classified each cluster based on the visual significance of its X-ray cavities: clusters were classified as having either convincing X-ray cavities, somewhat convincing X-ray cavities or unconvincing X-ray cavities. We then tabulated the classifications, and only kept those that were classified as having convincing or somewhat convincing X-ray cavities by at least 2 co-authors. This resulted in the rejection of 2 of the original 10 clusters: SPT-CLJ0334$-$4659 ($z=0.45$) and SPT-CLJ2043$-$5035 ($z=0.7234$). 

The final list of clusters with X-ray cavities is shown in Table \ref{tab1}. We note that 6 of these clusters were unanimously classified as having convincing or somewhat convincing X-ray cavities by all 3 co-authors. From now on, we refer to these as the clusters with ``convincing"  X-ray cavities. These are shown in the top portion of Table \ref{tab1}. The remaining 2 clusters, SPT-CLJ0156$-$5541 ($z=1.2$) and SPT-CLJ2342$-$5411 ($z=1.075$), were classified as having unconvincing X-ray cavities by 1 co-author. To emphasize the uncertainty related to these two particular systems, we highlight them differently in all tables (see bottom portion) and figures (square symbols instead of circles). 

The raw, unsharp-masked and King-subtracted images of the final list of clusters with X-ray cavities are shown in Fig. \ref{fig2}. The smoothing and binning factors are indicated in the lower-left corner of each image. ``BX" refers to the image binning factor: ``B1" means that the image was not binned, whereas ``B2" means that each pixel corresponds to 4 pixels in the original image. The smoothing factor is denoted as ``SX" where ``X" corresponds to the sigma of a Gaussian in units of pixels once the image was binned. For each unsharp-masked image, we show the 2 smoothing scales used to create the image, and for each King-subtracted image we show the smoothing scale adopted for the original image before creating and subtracting a King model. Those with brackets were additionally smoothed for illustrative purposes. The optical or near infrared images obtained as part of the SPT 2500 deg$^2$ follow-up campaign for photometric redshifts are also shown \citep{Hig2010723,Son2012761,Ble2014}. These were taken with the $Spitzer$ Space Telescope, the Swope 1-meter telescope, the Cerro Tololo Inter-American Observatory (CTIO) Blanco 4-meter telescope, or the Baade Magellan 6.5-meter telescope \citep[see][for details]{Ble2014}.

The depth of each cavity was estimated in the X-ray data using an azimuthal averaged surface brightness (see Fig.\ \ref{fig2b}). We measure azimuthal surface brightness profiles from the X-ray data (left panels of Fig.\ \ref{fig2}), using an annulus that encompassed the cavities, centered on the X-ray peak (middle-right panels of Fig.\ \ref{fig2}). Each annulus was divided into 8 azimuthal sectors, containing roughly $50$ counts per sector. On average, the depressions lie $\sim$1--3$\sigma$ below the surrounding X-ray emission and contain $10-40\%$ less counts, roughly consistent with X-ray cavities seen in local clusters. Based on these results, we refer to the cavities lying 1--2$\sigma$ below the surrounding X-ray emission as ``potential" cavities (see ``P" in Column 5 of Table \ref{tab1}), requiring deeper $Chandra$ data to confirm them, while we refer to those lying 2--3$\sigma$ (or more) below the surrounding X-ray emission as ``clear" cavities (see ``C" in Column 5 of Table \ref{tab1}). We note that the latter are only found among the 6 SPT-SZ clusters with visually ``convincing"  X-ray cavities, as explained earlier in this section. 

In Table \ref{tab1}, we also highlight the SPT-SZ clusters with X-ray cavities that have a radio source associated with the central regions. Since all of the low-redshift clusters with X-ray cavities have a radio-active BCG, we expect this to be similarly the case for the SPT-SZ clusters with X-ray cavities. We use the 843 MHz Sydney University Molonglo Sky Survey \citep[SUMSS, synthesized beam width of $\sim$40$^{\prime\prime}$;][]{Boc1999117,Mau2003342}, but owing to the large beam size, we cannot resolve any extended jet-like emission. We therefore use the radio data simply to verify that the clusters in our sample with X-ray cavities have radio-loud BCGs. We note that the radio sources in SPT-CLJ0033$-$6326 and SPT-CLJ0156$-$5541 are not centered on the BCG, but that the BCG lies well within the beam size of SUMSS. While the position uncertainty of sources detected in SUMSS is only $\sim$10$^{\prime\prime}$, these BCGs may still contain a radio source that is contributing to the overall emission in the beam, since they lie well within the beam size. We therefore consider that these sources have a radio counterpart for the purposes of this paper. SPT-CLJ0509$-$5342 lies within two beam sizes of a nearby 120 mJy radio source, making its detection by SUMSS uncertain since SUMSS is dynamic range limited by 1:100 on average \citep{Boc1999117}. This bright radio source is most likely not associated with the BCG since, at its redshift, it would be located some 700 kpc from the galaxy. Finally, we find no evidence from the SUMSS maps of a radio source in SPT-CLJ2342$-$5411. Considering that this source is the second most distant cluster among our candidates ($z=1.075$), the non-detection may simply be due to the limited sensitivity of the survey. In summary, and as expected, the majority of clusters with clear X-ray cavities (Table \ref{tab1}) have a detected radio counterpart.

\begin{figure}
\centering
\begin{minipage}[c]{0.99\linewidth}
\centering \includegraphics[width=\linewidth]{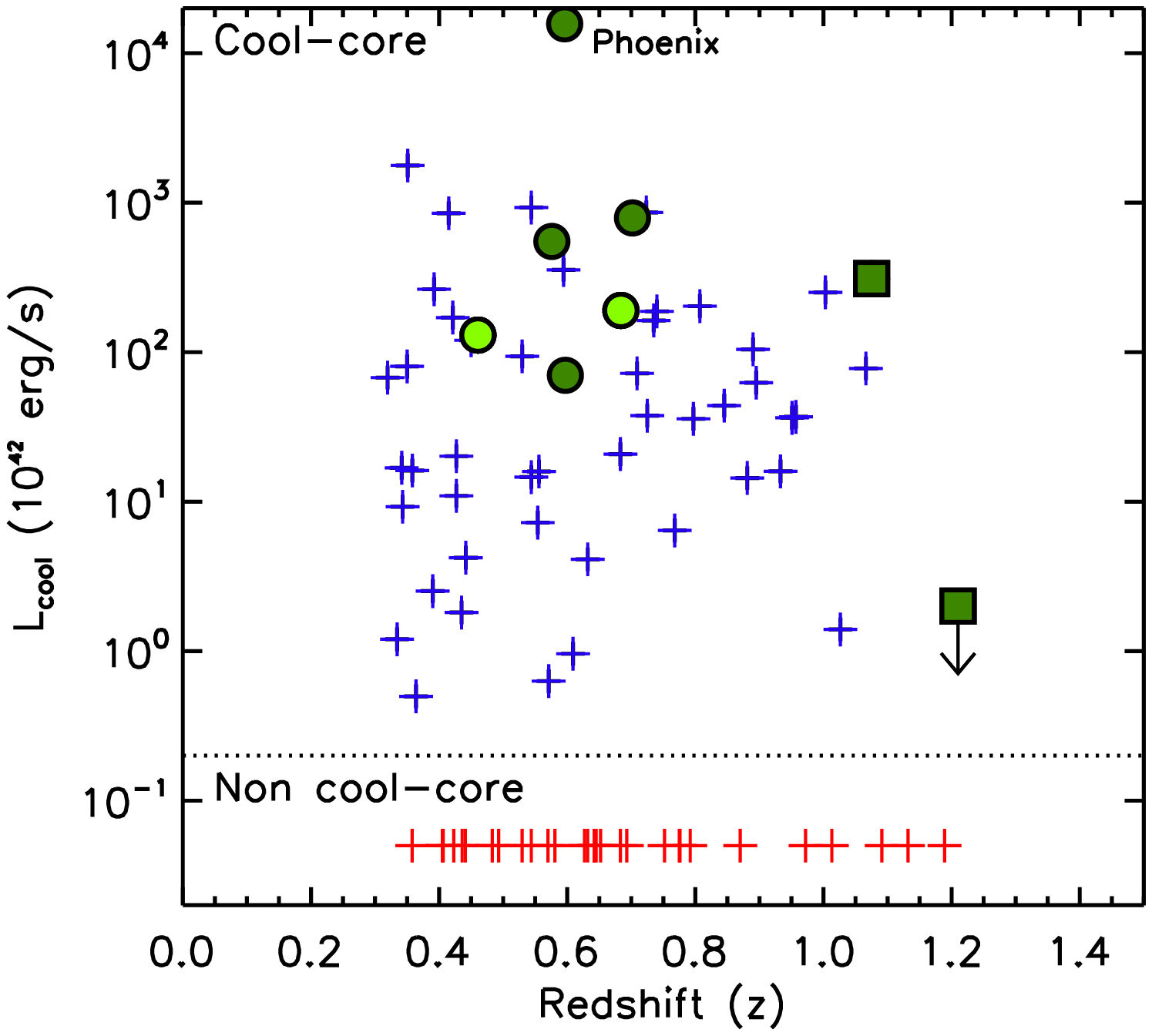}
\end{minipage}
\caption[]{Cooling luminosities ($L_{\rm cool}$) at 7.7 Gyrs as a function of redshift for the 83 SPT-SZ clusters with $Chandra$ X-ray observations. The cool cores, where the cooling time drops below 7.7 Gyrs, are shown in blue and the non-cool cores in red. For the latter, we assign an arbitrary low $L_{\rm cool}$ value below the dotted line for illustrative purposes. Highlighted in green are the systems in which we identified candidate X-ray cavities. The different shades of green hilight the depth of the cavities compared to the surrounding X-ray emission with light (dark) green for the ``clear" (``potential") cavities. The clusters with convincing cavities are shown with the circles, while SPT-CLJ0156$-$5541 and SPT-CLJ2342$-$5411 are shown with the squares.}
\label{fig3}
\end{figure}
\subsection{Cavity energetics}

To study the evolution of AGN feedback in BCGs, the energetics of the X-ray cavities in the SPT-SZ sample must be computed. These are estimated using the standard techniques, which we describe below \citep[see][and references therein]{Bir2004607}. Assuming that the cavity is filled with a relativistic fluid, the total enthalpy is: 
\begin{equation}
E_{\rm bubble}= 4pV \, ,
\label{eq2}
\end{equation}
where $p$ is the thermal pressure of the ICM at the radius of the bubble (estimated from X-ray data, assuming $p = n_ekT$), and $V$ is the volume of the cavity. We assume that the cavities are of prolate shape. The volume is then given by $V=4{\pi}R^2_{\rm w}R_{\rm l}/3$, where $R_{\rm l}$ is the projected semi-major axis along the direction of the jet, and $R_{\rm w}$ is the projected semi-major axis perpendicular to the direction of the jet. Errors on the radii are assumed to be $\pm$20$\%$, and the jet is defined as the line that connects the central AGN to the middle of the cavity. The position of the central AGN is chosen to coincide with the position of the BCG as seen from the optical or infrared images (Fig. \ref{fig2}). If two central dominant galaxies were present in the optical images, we chose the brightest one as the BCG. There is only one cluster where this applies: SPT-CLJ0616$-$5227. Modifying the location of the BCG to coincide with the second dominant galaxies modifies the cavity energetics by a factor of $\leq2$, which is not significant for the purposes of this study. 

In Table \ref{tab1}, we give the constraints on the X-ray cavity radii ($R_{\rm l}$ and $R_{\rm w}$), enthalpy ($pV$), and cavity power ($P_{\rm cav}$). Cavity powers are determined by dividing the total enthalpy of the X-ray cavity ($4pV$) with its age. The latter is given by the buoyancy rise time \citep[][]{Chu2001554}:
\begin{equation}
t_\mathrm{buoyancy}= R\sqrt{\frac{SC_{\rm D}}{2gV}} \, .
\label{eq3}
\end{equation}
Here, $R$ is the projected distance from the central AGN to the middle of the cavity, $S$ is the cross-sectional area of the bubble ($S={\pi}R_{\rm w}^2$), $C_{\rm D}$ is the drag coefficient \citep[assumed to be 0.75;][]{Chu2001554}, and $g$ is the local gravitational potential. We use the values of $g$ derived by M13 (see resulting values in Table \ref{tab1}), but stress that the typical uncertainties are larger than those reported in M13: they are on the order of $\pm50\%$ as mentioned in Section 2. The enthalpy and powers of the X-ray cavities in the SPT-SZ sample are therefore known, at best, to within a factor of a few.

\begin{figure}
\centering
\begin{minipage}[c]{0.99\linewidth}
\centering \includegraphics[width=\linewidth]{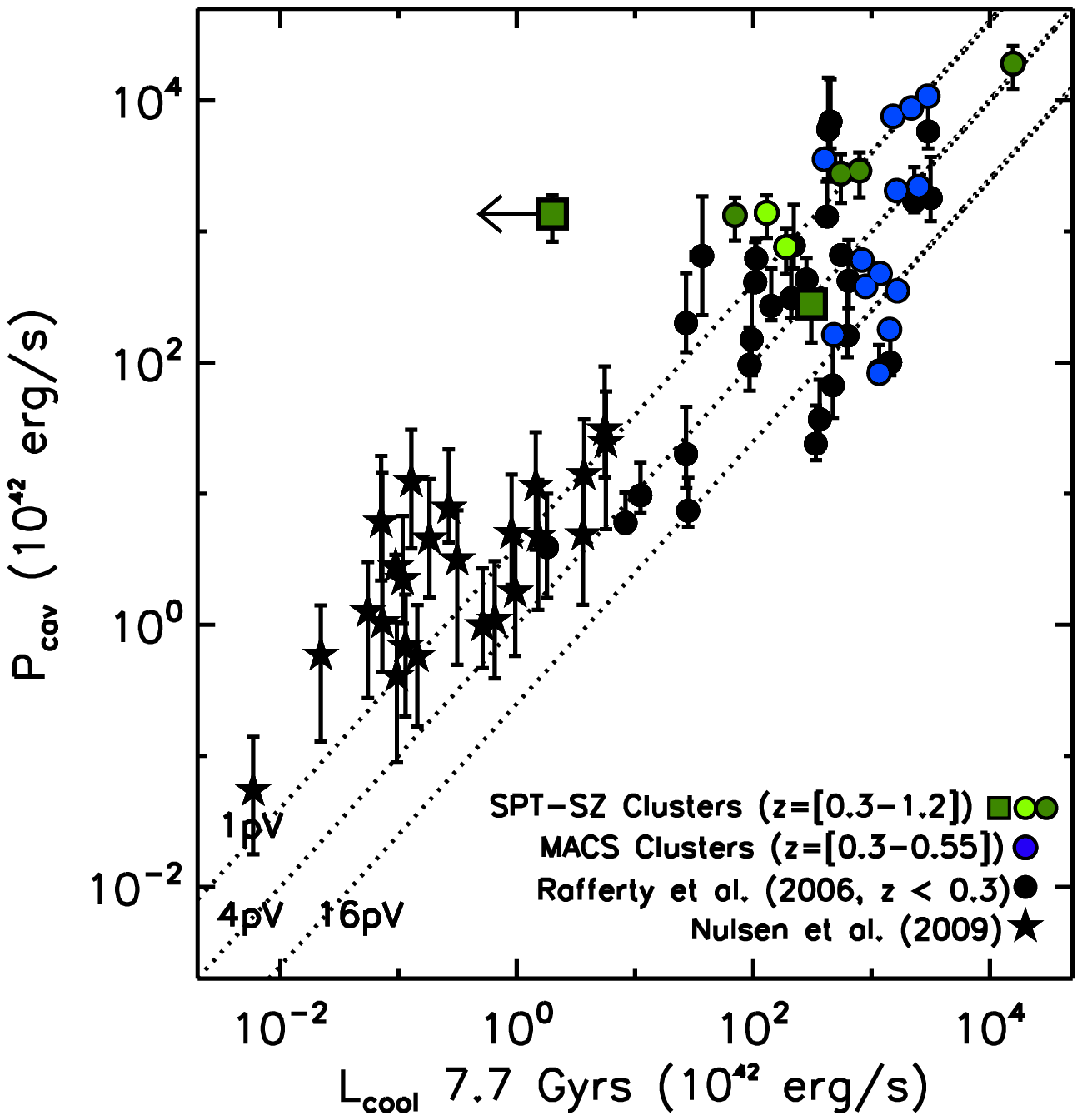}
\end{minipage}
\caption[]{Comparison between the mechanical power being injected by the AGN in the BCG ($P_{\rm cav}$) and the cooling luminosity ($L_{\rm cool}$) of the cluster at 7.7 Gyrs. The SPT-SZ clusters with candidate X-ray cavities are shown in green: circles are used to hilight the 6 SPT-SZ clusters with convincing X-ray cavities, while squares are used for SPT-CLJ0156$-$5541 and SPT-CLJ2342$-$5411. The different shades of green hilight the depth of the cavity compared to the surrounding X-ray emission with light (dark) green for the ``clear" (``potential") cavities. We note that the $P_{\rm cav}$ values are uncertain by a factor of a few for the SPT-SZ clusters as discussed in Section 2.}
\label{fig4}
\end{figure}

\section{Cooling luminosities}
Ultimately, we wish to determine if AGN feedback is operating differently in high-redshift clusters and, in particular, if AGN feedback is able to suppress cooling of the hot ICM at early times. We can address this question by comparing the mechanical energy injection (measured from X-ray cavities) to the energy lost by cooling. The latter is often quoted in terms of the cooling luminosity ($L_{\rm cool}$), defined as the bolometric X-ray luminosity ($0.01-100\keV$) interior to the radius where the cooling time is equal to some threshold value. We adopt several definitions of this value throughout this paper.

The first definition is motivated by earlier studies on X-ray cavities, including those by \nocite{Nul2009,Raf2006652}Nulsen et al. (2009), Rafferty et al. (2006) and HL12. These studies defined the cooling luminosity to be the bolometric X-ray luminosity within which the cooling time is equal to 7.7 Gyrs (Column 3 in Table \ref{tab1}). For the clusters in this sample, this definition effectively measures the total cooling occurring within a radius of $\lesssim100$ kpc (see Table \ref{tab1}). As in HL12, we define the cooling time to be:

\begin{equation}
t_{\rm cool} = \frac{5}{2}\frac{n~kT~V}{L_{\rm X}} \,  .
\label{eq4}
\end{equation}

Here, $n$ is the total number of gas particles per unit volume, $kT$ is the gas temperature, $L_{\rm X}$ is the gas bolometric X-ray luminosity and $V$ is the gas volume contained within each annulus. In addition to the internal energy of the gas ($3/2\times{n}{kT}$), the enthalpy includes the work done on the gas as it cools at constant pressure in Eq. \ref{eq4} ($5/2\times{n}{kT}$). \citet{Nul2009} calculate their cooling luminosities within a radius where the cooling time is 7.7 Gyrs, but they do not specify if these luminosities are bolometric luminosities. Since their sample contains only lower-mass systems, we only compare our study to theirs qualitatively. \citet{Raf2006652} calculate their bolometric cooling luminosities at 7.7 Gyrs, but they do not specify the equation they use for the cooling time. In HL13, we recalculated the cooling luminosities using Eq. \ref{eq4} for all the massive clusters in \citet[][14 in total]{Raf2006652}. We found our values to be consistent with theirs within 1$\sigma$ of the scatter in the population, although ours were systematically smaller. Overall, there is no significant difference, at least for the purposes of this study, and we proceed to directly compare our results with \citet{Raf2006652}. 

The second definition, chosen to be the bolometric X-ray luminosity within $r=50$ kpc (Column 4 of Table \ref{tab1}), is more physically motivated. This definition allows us to directly compare the heating and cooling within (roughly) the same volume. The values we obtain with this second definition are $20-40\%$ smaller compared to those obtained at 7.7 Gyrs. 

Finally, we adopt a third definition, motivated by the results of M13 (Column 5 of Table \ref{tab1}). M13 estimated that cool cores began to assemble in massive systems at $z=1^{+1.0}_{-0.2}$, implying that only a small fraction of the ICM in high-redshift clusters would have had the time to cool completely, form cold molecular gas and feed the black hole that is generating the X-ray cavities. As such, we define a third cooling luminosity to be the bolometric X-ray luminosity within which the cooling time is equal to the look-back time since $z=2$. The new values with this third definition are $10-95\%$ smaller compared to those obtained at 7.7 Gyrs. For the 13 MACS clusters with X-ray cavities (HL12), we also computed cooling luminosities using this third definition and find that the new values are only $\sim10\%$ smaller. It therefore remains appropriate to use the definition of the cooling luminosity at $7.7$ Gyrs for MACS clusters ($z_{\rm average}\sim0.4$). The difference only becomes significant for higher redshift clusters, such as those presented in this work ($z_{\rm average}\sim0.7$).

\begin{figure*}
\centering
\begin{minipage}[c]{0.45\linewidth}
\centering \includegraphics[width=\linewidth]{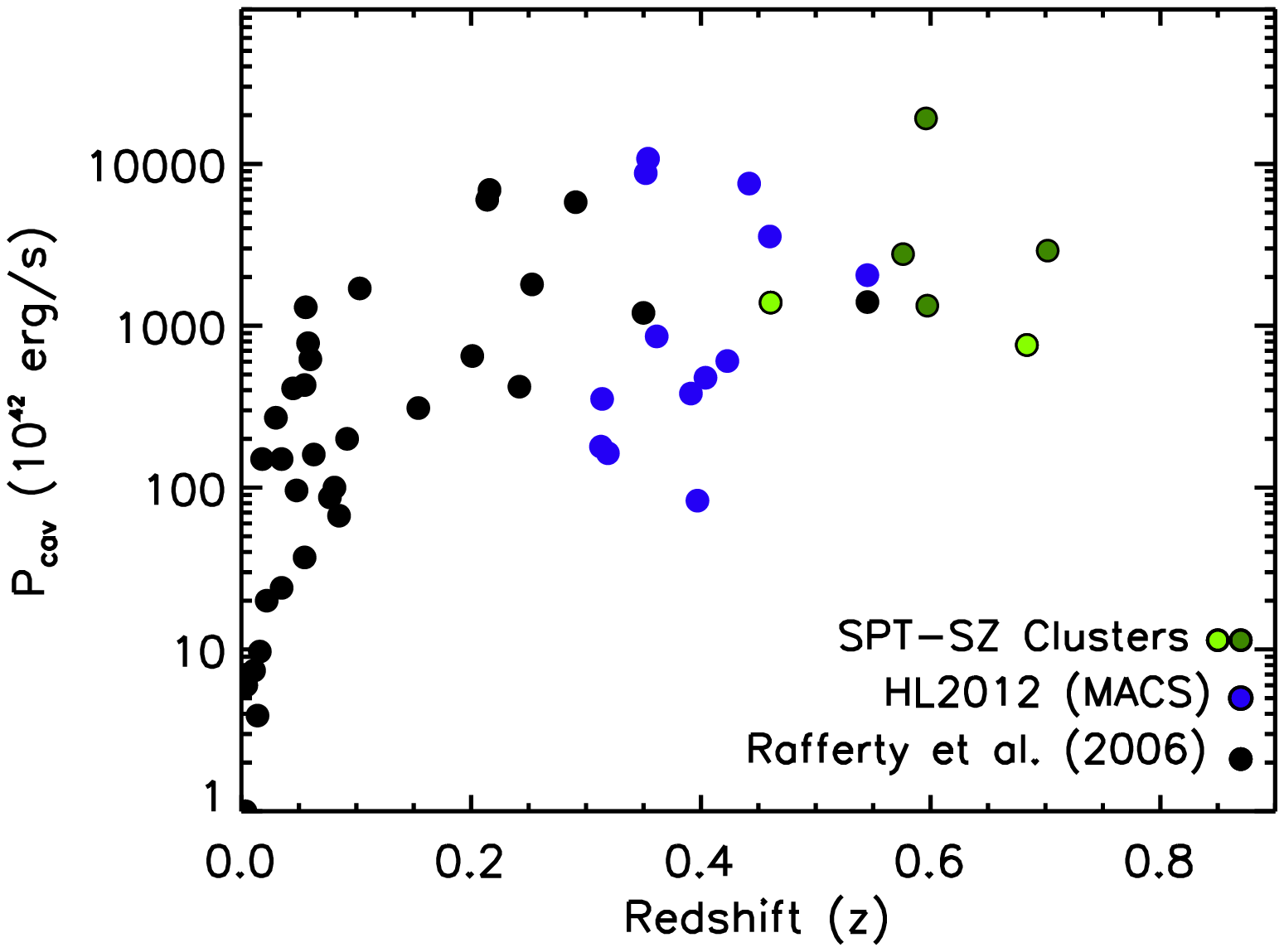}
\end{minipage}
\begin{minipage}[c]{0.45\linewidth}
\centering \includegraphics[width=\linewidth]{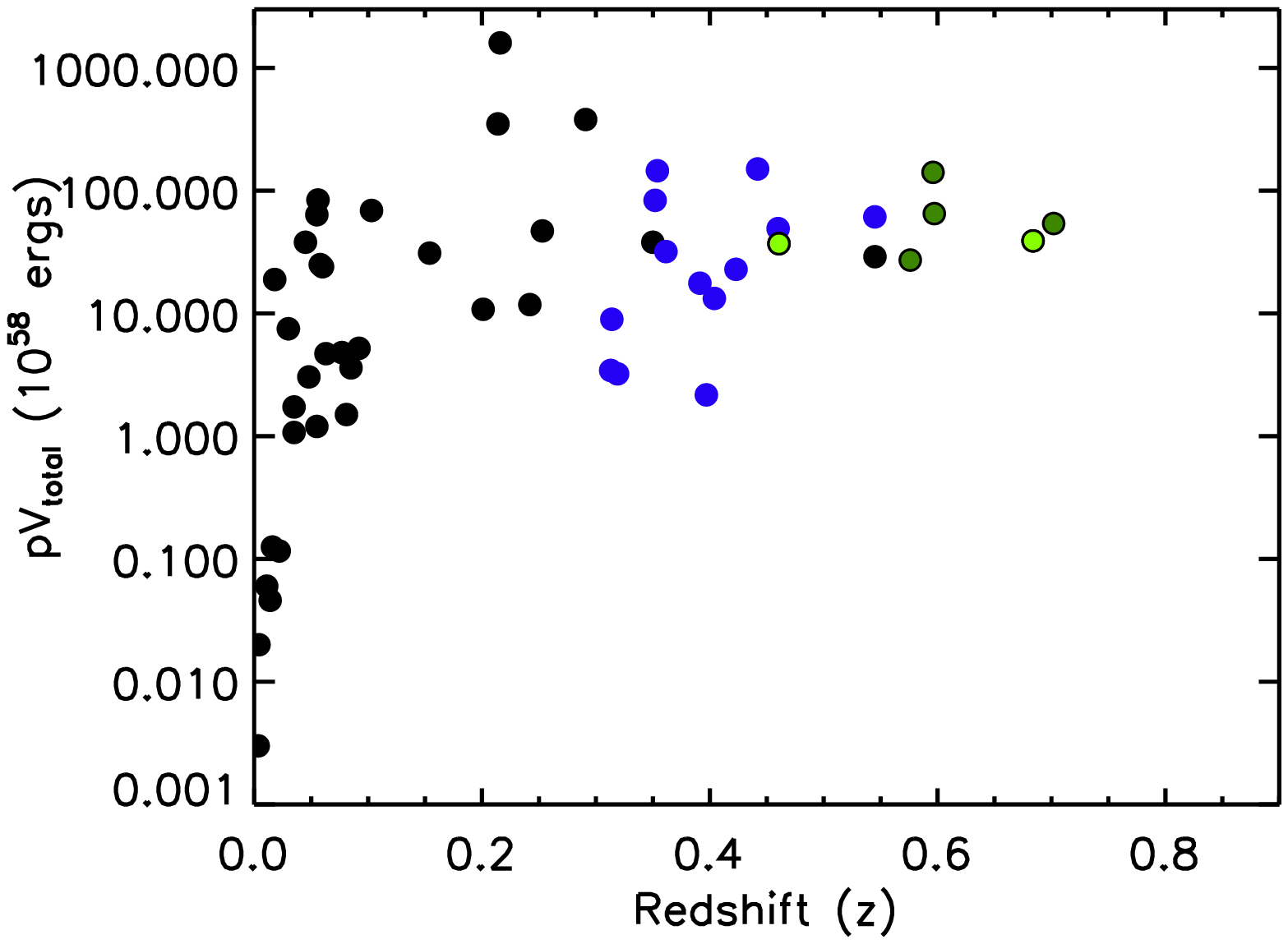}
\end{minipage}
\begin{minipage}[c]{0.45\linewidth}
\centering \includegraphics[width=\linewidth]{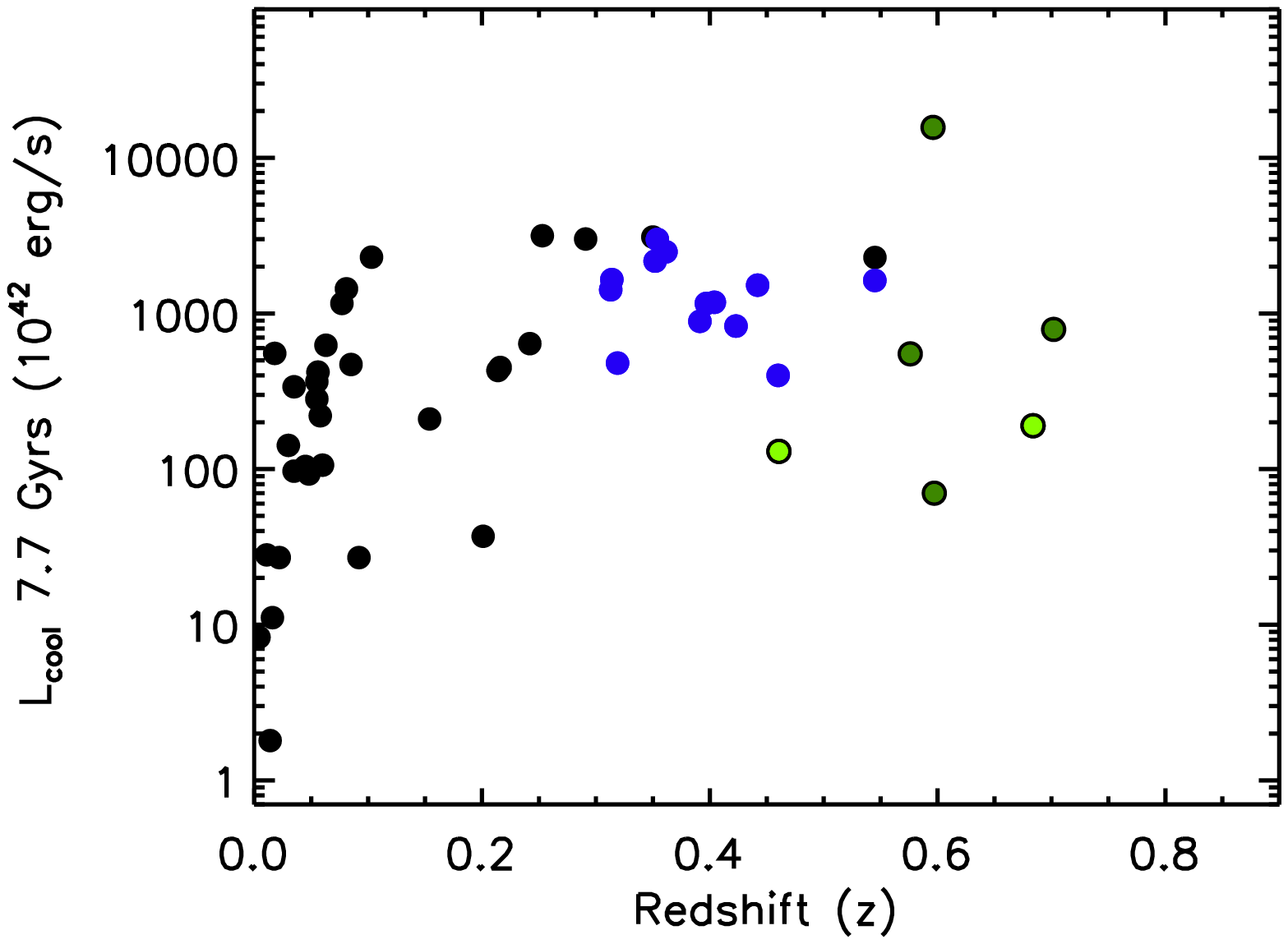}
\end{minipage}
\begin{minipage}[c]{0.45\linewidth}
\centering \includegraphics[width=\linewidth]{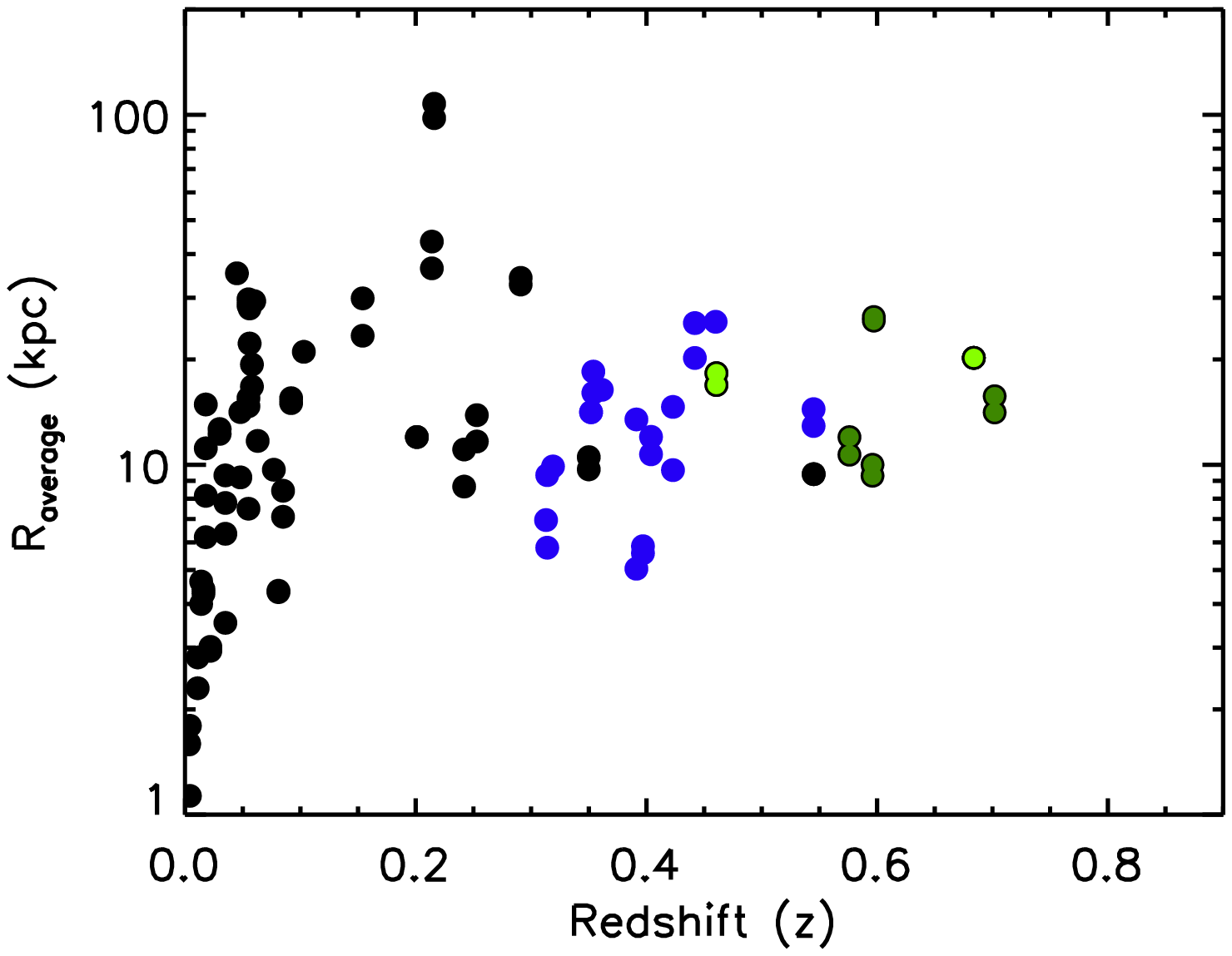}
\end{minipage}
\caption[]{Plots of the mechanical power of X-ray cavities ($P_{\rm cav}$, top-left), the enthalpy of the X-ray cavities ($PV_{\rm total}$, top-right), and the cooling luminosity as defined with the 7.7 Gyrs threshold ($L_{\rm cool}$, bottom-left) for each cluster, as a function of redshift. Note that the $P_{\rm cav}$ and $PV_{\rm total}$ values are usually uncertain by at least a factor of a few, especially for SPT-SZ clusters as discussed in Section 2. We also show the average radius of $each$ cavity ($R_{\rm average}$, bottom-right) as a function of redshift. Same symbols as Fig. \ref{fig4}. Here, we focus only on the SPT-SZ clusters with visually convincing cavities, and therefore exclude SPT-CLJ0156$-$5541 and SPT-CLJ2342$-$5411 from these plots. }
\label{fig5}
\end{figure*}

\section{Results}

\begin{figure}
\centering
\begin{minipage}[c]{0.99\linewidth}
\centering \includegraphics[width=\linewidth]{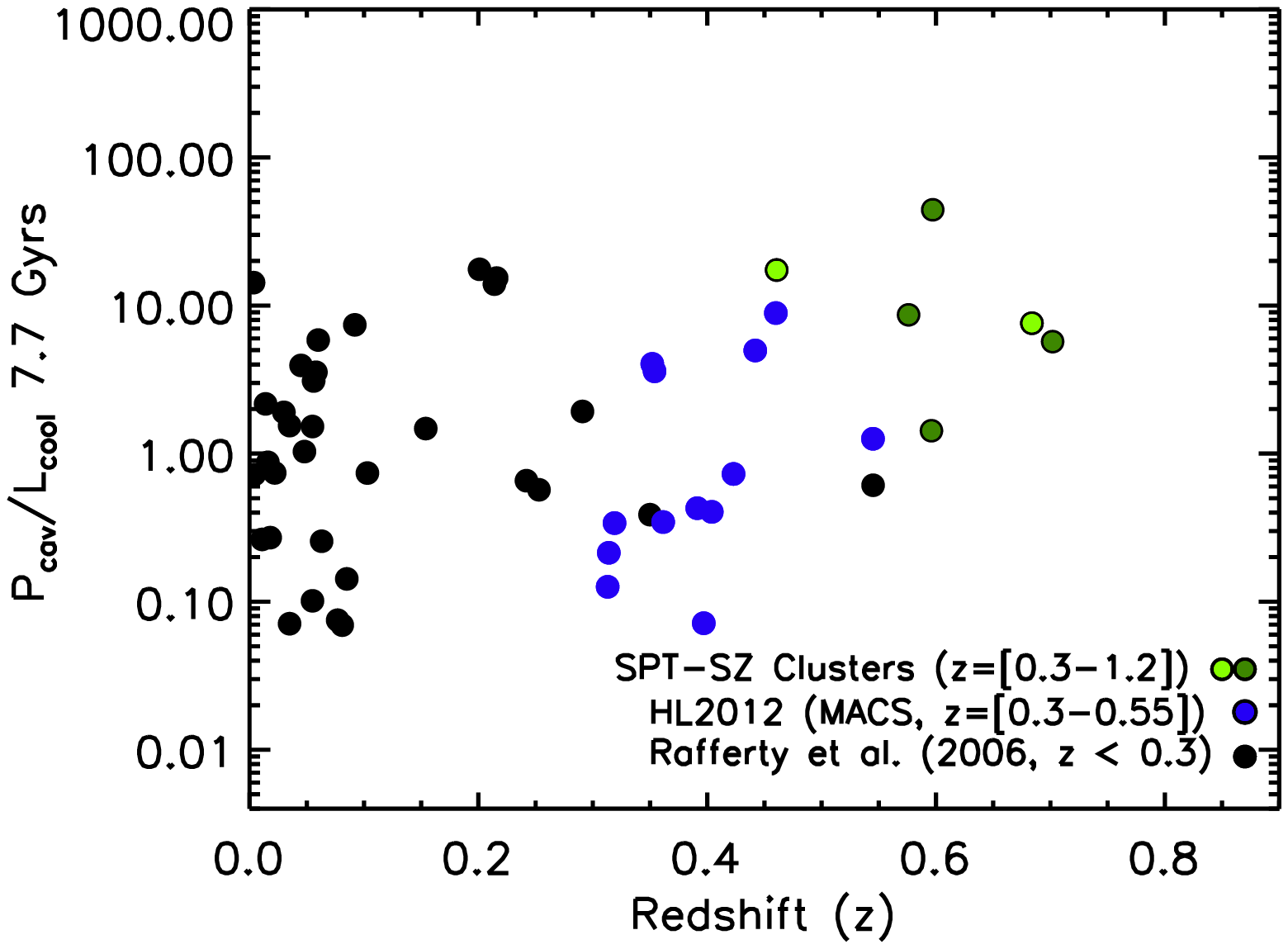}
\end{minipage}
\caption[]{Plot showing the ratio between the mechanical power of X-ray cavities ($P_{\rm cav}$) and the cooling luminosity ($L_{\rm cool}$) at 7.7. Gyrs for different samples of clusters, as a function of redshift. We also exclude SPT-CLJ0156$-$5541 and SPT-CLJ2342$-$5411 from these plots, as in Fig. \ref{fig5}. }
\label{fig6}
\end{figure}

\subsection{Distribution of X-ray cavities}
In Fig. \ref{fig3}, we plot the cooling luminosity of the 83 SPT-SZ clusters with $Chandra$ observations using the standard 7.7 Gyrs cooling time definition. This plot shows that the majority of the SPT-SZ clusters with X-ray cavities lie in the strongest cool-core clusters. This is expected in part since the majority of the counts in the $Chandra$ X-ray images will be concentrated toward the central regions for cool-core clusters, due to highly-peaked X-ray surface brightness distributions. This higher concentration of counts in the central regions makes it easier to identify depressions.

\subsection{Energetics of X-ray cavities}
AGN feedback in clusters of galaxies is known to be finely tuned to the energy needed to offset cooling of the hot ICM out to $z\sim0.6$ (HL12). We illustrate this in Fig. \ref{fig4}, where we plot data points from \citet[][black circles]{Raf2006652}, \citet[][black stars]{Nul2009} and HL12 (blue circles). In green, we show the SPT-SZ clusters with X-ray cavities. Here, we adopt the first definition of the cooling luminosity, defined as the bolometric X-ray luminosity within which the cooling time is equal to 7.7 Gyrs. Fig.\ \ref{fig4} shows that, on average, X-ray cavities can offset cooling of the hot ICM even in the highest redshift sources. We note that this result remains true if we use the other two definitions of the cooling luminosity (see Section 4). We discuss these results in Section 6.4.1.

\subsection{Redshift distribution of X-ray cavities}
In Fig.\ \ref{fig5}, we address whether AGN feedback in clusters of galaxies, as probed by the presence and properties of X-ray cavities, is evolving significantly with redshift. In this figure, we plot the mechanical power of the X-ray cavities ($P_{\rm cav}$), the enthalpy of the X-ray cavities ($PV_{\rm total}$), and the cooling luminosity defined with the 7.7 Gyrs threshold ($L_{\rm cool}$) for each cluster, as a function of redshift. In the bottom-right panel, we also plot the average X-ray cavity radius $R_{\rm average}$ defined as $(R_l{\times}R_w)^{1/2}$ for $each$ X-ray cavity. Since we plot the radius for $each$ cavity, there are more data points in this panel. We only include the objects in \citet[][black circles]{Raf2006652} and HL12 (blue circles) since \citet[][]{Nul2009} do not provide any redshift information. In these plots and for the remainder of the paper, we focus only on the systems with convincing cavities (i.e. we exclude SPT-CLJ0156$-$5541 and SPT-CLJ2342$-$5411). Fig. \ref{fig5} shows that there is no significant evolution in any of the cavity properties for the largest and most powerful outflows. To provide a first order correction for the dependency of mass on the cavity powers, we also plot in Fig. \ref{fig6} the ratio between the mechanical power and the cooling luminosity as a function of redshift. As we will discuss in Section 6.1, we are strongly biased against finding small cavities ($r\lesssim10$ kpc) in the XVP $Chandra$ data. Removing these small cavities from Fig. \ref{fig6} for the lower-redshift samples does not affect the results. We further discuss Figs. \ref{fig5} and \ref{fig6} in Section 6.4.1.

\section{Discussion}
We have visually inspected the $Chandra$ X-ray images of a sample of 83 SPT-SZ clusters located at $0.3<z<1.2$, and found that 6 clusters contain visually convincing surface brightness depressions that we interpret to be X-ray cavities. By comparing the X-ray emission within the cavities to the surrounding X-ray emission, we determined the depth of the cavities in each system and classified them into two distinct categories: those with ``potential" cavities (4/6 clusters) when the cavity emission was only $1-2\sigma$ below the surrounding X-ray emission and those with ``clear" cavities (2/6) when the cavity emission was at least $2-3\sigma$ below the surrounding X-ray emission. While deeper $Chandra$ observations are needed to confirm the ``potential" cavities, we proceed with the discussion of these results. We first discuss the various selection effects that may be present in this study, and then we discuss the implications of this study. 

\subsection{Selection effects}

\begin{figure}
\centering
\begin{minipage}[c]{0.99\linewidth}
\centering \includegraphics[width=\linewidth]{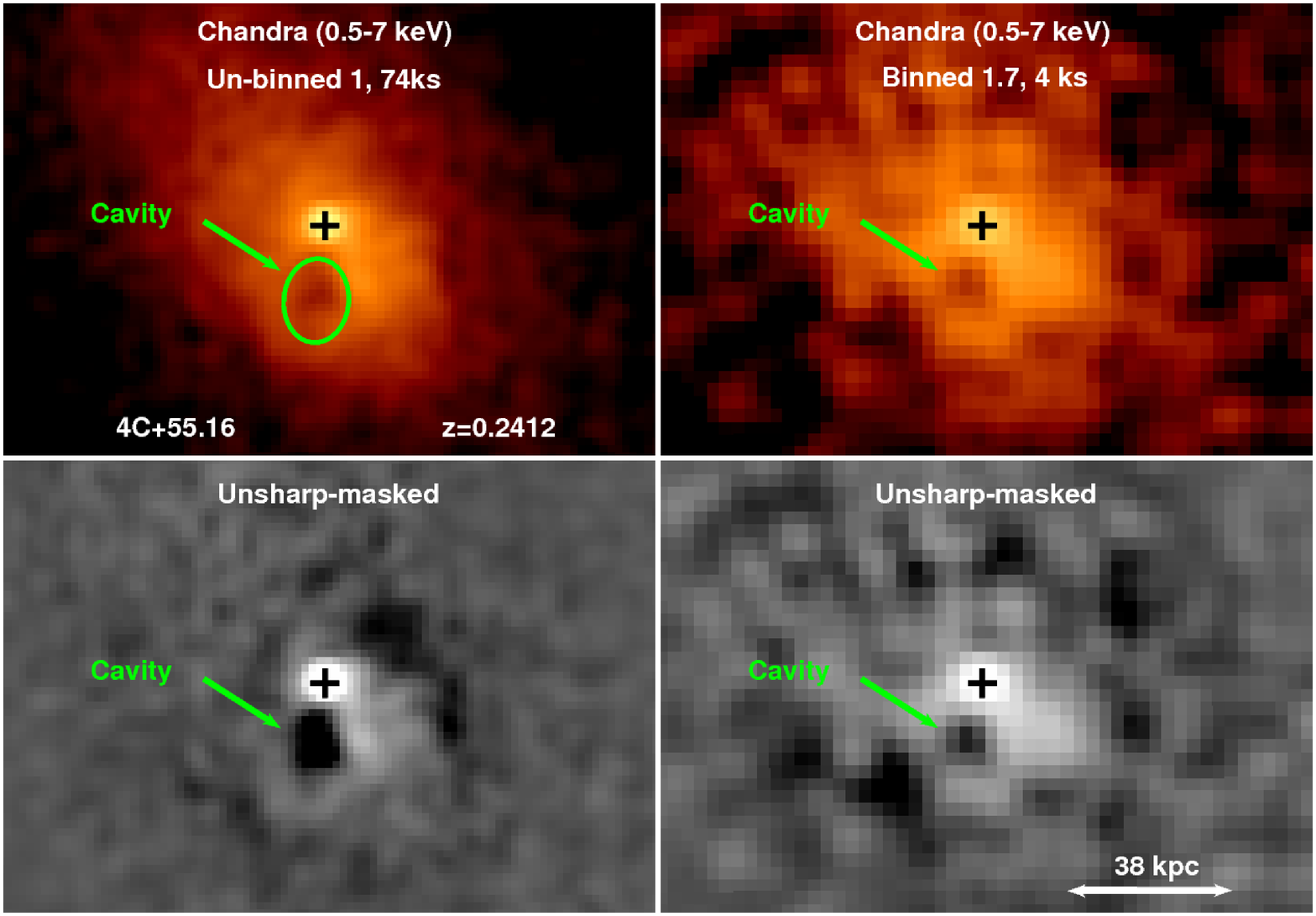}
\end{minipage}
\caption[]{$Chandra$ X-ray images of 4C+55.16 ($z=0.2412$). Left: Un-binned image (74 ks, $\sim40000$ counts, top) and corresponding unsharp-masked image (bottom). Right: Limited exposure image (4 ks, $\sim2000$ counts, top) binned by a factor of 1.7 to mimic its appearance as if it were at the average redshift of the SPT-SZ clusters with X-ray cavities ($z\sim0.7$), and corresponding unsharp-masked image (bottom). This figure shows that the southern X-ray cavity remains visible in all X-ray images.}
\label{fig8}
\end{figure}
\begin{figure}
\centering
\begin{minipage}[c]{0.8\linewidth}
\centering \includegraphics[width=\linewidth]{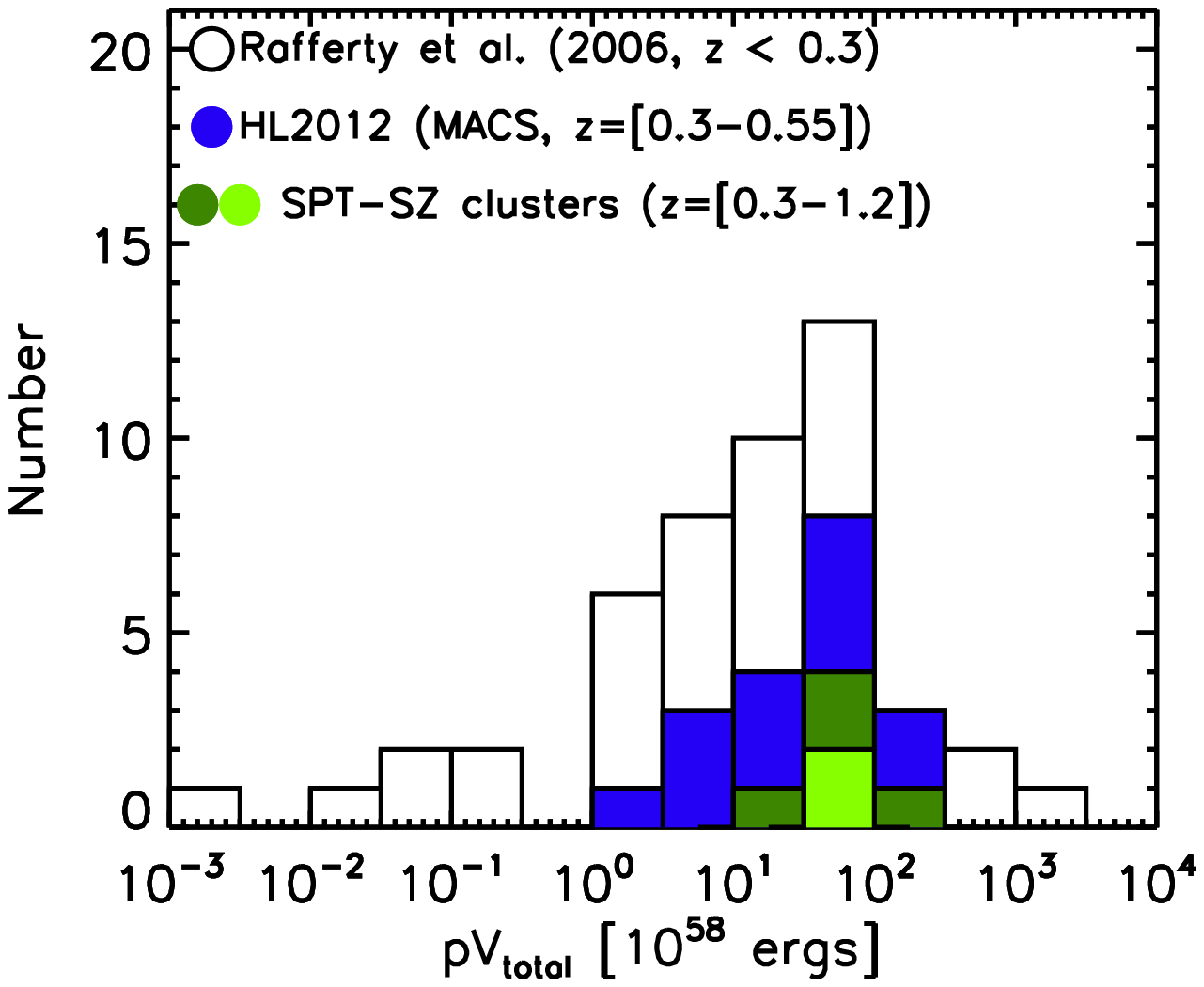}
\end{minipage}
\begin{minipage}[c]{0.8\linewidth}
\centering \includegraphics[width=\linewidth]{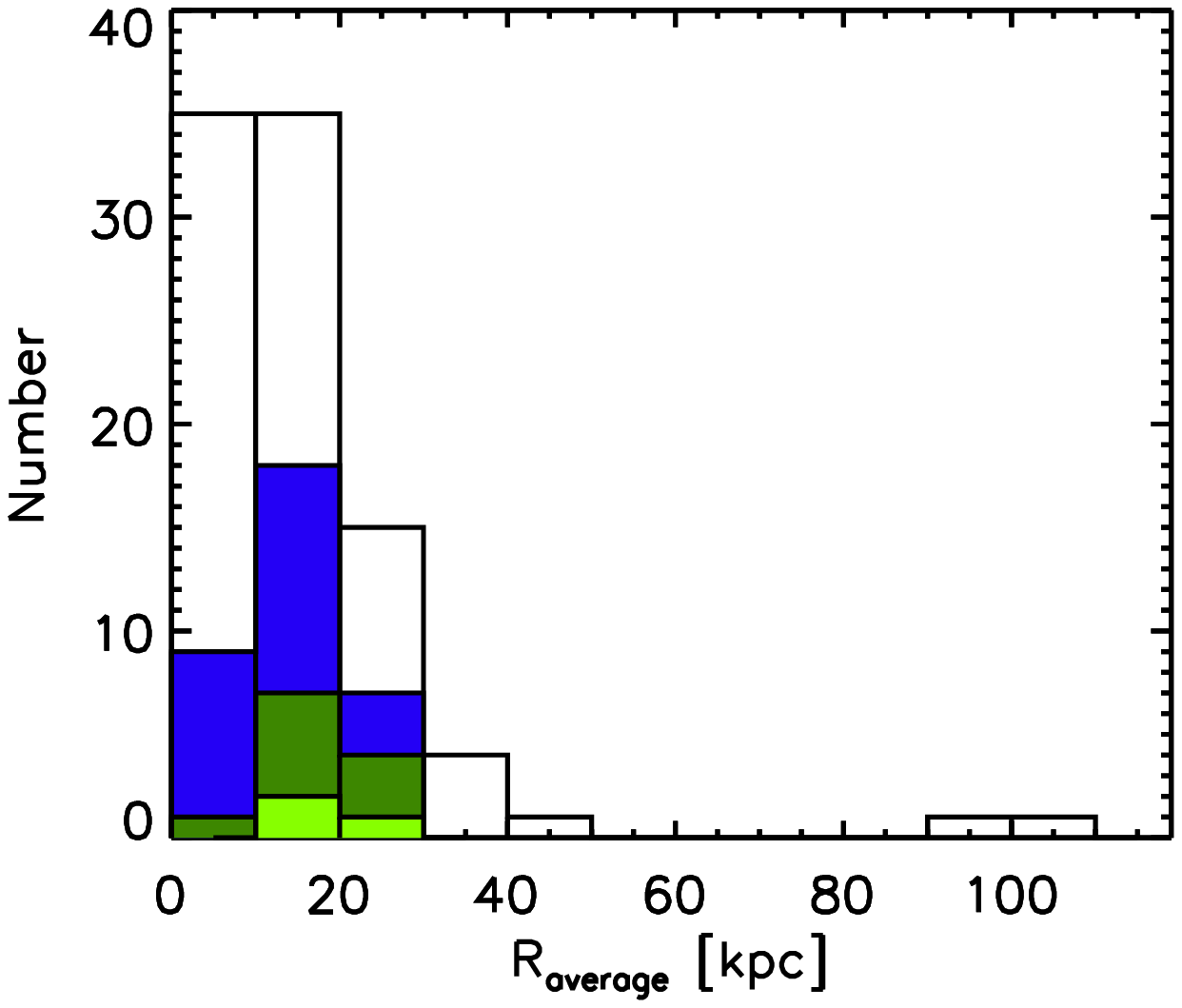}
\end{minipage}
\caption[]{Plots hilighting the distribution of cavity enthalpy (top) and the average cavity radius (bottom). As in Figs. \ref{fig5} and \ref{fig6}, we exclude SPT-CLJ0156$-$5541 and SPT-CLJ2342$-$5411 from these plots.}
\label{fig7}
\end{figure}

While the results presented in Section 5 are interesting, there may be several biases present in the cavity selection method and the analysis. Below we address these potential biases. 

First, we will clearly miss cavities below the resolution limit ($R_{\rm average}~\aplt~10$ kpc, see Table \ref{tab1}), as well as those that lie on a jet axis parallel to our line of sight. Furthermore, we limited the search to the central regions of each cluster, where most of the X-ray counts are located, and disregarded any X-ray depression identified at $r>100$ kpc. We are likely therefore to miss any extremely large cavities like those in MS 0735.6+7421, but note that these are expected to be rare with only two such systems known so far \citep{McN200745}. 

In Fig. \ref{fig8}, we show the $Chandra$ X-ray images of the massive galaxy cluster 4C+55.16 \citep[$z=0.2412$,][]{Hla2011415,Iwa2001328}. The left panels show the deep $Chandra$ images (74 ks), which reveal a large southern X-ray cavity. The right panels show the same image but for a reduced exposure time of only 4 ks and binned by a factor of 1.7 (pixel size of $0.836''\times0.836''$) to mimic the appearance of the cluster as if it were located at the average redshift of the SPT-SZ clusters with X-ray cavities ($z\sim0.7$). We reduce the exposure time to 4 ks since this reduces the total number of cluster counts to $\sim2000$, similar to the counts in the $Chandra$ X-ray images of the SPT-SZ clusters. Fig. \ref{fig8} shows that the cavity in 4C+55.16 remains visible in all panels. According to our criteria described in Section 3.1, we would identify this X-ray cavity as a ``potential" cavity since the cavity emission lies only $1-2\sigma$ below the surrounding X-ray emission. We note that adding an additional $75$ks blank field exposure to the images, mimicing the increase in background from the long exposure needed to get 2000 counts for high-redshift clusters, does not change our results. 

Even though Fig. \ref{fig8} demonstrates that we can identify cavities with as few as 2000 counts, we are likely missing several cavities due to limited data quality of the 83 SPT-SZ clusters with $Chandra$ observations. To test the probability of identifying cavities in the SPT-SZ cluster survey, we applied the same imaging processes as in Fig. \ref{fig8} to the 13 MACS clusters with X-ray cavities: we reduced the exposure times such that each cluster contained only $\sim$2000 counts and binned the images to mimic the appearance of each cluster as if it were located at $z\sim0.7$. Interestingly, we found that only $\sim$60\% of these X-ray cavities would then have been detected using the same criteria as those in Section 3.1, and that an even smaller fraction ($\sim$20\%) would have had ``clear" cavities with a cavity emission $2-3\sigma$ (or more) below the surrounding X-ray emission, as defined in Section 3.1. This demonstrates that we are likely missing cavities in the SPT-SZ sample due to the limited X-ray depth of the survey \citep[see also][]{En2002384,Die2008687,Bir2012427}. Moreover, the strongly peaked X-ray surface brightness distributions of strong cool-core clusters also implies that the X-ray counts will be highly concentrated towards the central regions in these clusters. We may therefore be preferentially selecting X-ray cavities in strong cool-core clusters. 

Simulations have shown that the increased SZ signal due to the presence of a cool core is not significant \citep[][]{Mot2005623,Pip2010404}, and that, similarly, star formation does not fill in the SZ decrement significantly \citep[e.g.,][]{McD2012Nat}. In particular, \citet[][]{Lin2009694} showed that at $z=0.6$, less than 2$\%$ of clusters with a mass exceeding $10^{14}M_{\rm \odot}$ will host a radio AGN that contributes to more than 20$\%$ of the SZ signal, even when accounting for the fraction of systems with flat or inverted spectra. This fraction is even smaller at higher redshift due to the decreasing flux. These biases should therefore not be significant. 

Finally, we recall that the majority of the cavities presented here lie only $1-2\sigma$ below the surrounding X-ray emission. These depressions could therefore be caused by other activity in the cluster such as merger-induced asymmetries in the X-ray gas distribution. This is especially true for SPT-SZ clusters, since cluster merger rates are expected to increase with redshift \citep[e.g.,][]{Coh200524,Ett2008387}. It is also important to note that cluster mergers can provide an additional source of heating to the X-ray gas. At high redshift, AGN feedback may therefore no longer be the only heating source of the ICM.

\subsection{X-ray cavity detection}

Out of the 83 SPT-SZ clusters with $Chandra$ observations, we find that 6 clusters have visually convincing X-ray cavities. For these 6 systems, all of the cavities are found in pairs, most of which are similar in size and symmetrically located on either side of the AGN in the BCG. In Section 3.1, we also showed that these cavities lie $1-3\sigma$ below the surrounding X-ray emission. While the probability of having one $1-3\sigma$ random fluctuation in the $Chandra$ X-ray images is high, due to the Poisson nature of the observations, the probability of having two such fluctuations in an 8-element azimuthal array, as in Fig. \ref{fig2b}, is substantially lower: $\sim$37\% for two $\sim$1$\sigma$ fluctuations within $r=100$ kpc. Moreover, the probability of having the two fluctuations at $180\pm30$ degrees from one another in PA is only $\sim$10\%. If we also require them to have matching radii to within $\pm50$ kpc in radius from one another, as observed in Table \ref{tab1}, then the number drops to $\sim$7.5\% and even further for two $\sim$2$\sigma$ ($\lesssim$0.2\%) and two $\sim$3$\sigma$ ($\lesssim$0.001\%) fluctuations. Overall, the probability that the cavities presented here are caused by random fluctuations, given the PA and radii properties observed in Table \ref{tab1}, is very small. Instead, the depressions may well be cavities being carved out by twin radio lobes. 

We note that all of these calculations were computed using a Monte Carlo approach, assuming a random normal distribution for the cavity flux and a random uniform distribution for the PAs and radii of the cavities. The statistics are also based on the annuli shown in Fig. \ref{fig2b}, which do not cover the full radial extent ($r<100$kpc) considered in the cavity selection. By adding additional spatial elements we would increase the chance of a false detection (by adding more draws), but would also increase the significance of the individual detections (by improving the measurement of the ``background''). Thus, we expect that these probabilities are within a factor of $\sim$2 of what one would get by doing a more rigorous analysis. In principle, it should also be possible to derive an approach, similar to the one used here, but focusing on the probability of finding $1-3\sigma$ fluctuations by eye within the entire region of $r<100$ kpc. However, it is very difficult to quantify this selection function, as visual identification can vary significantly from one person to another. We therefore choose to focus only on reporting approximate probabilities in this paper, based on the annuli statistics shown in Fig. \ref{fig2b}.

\subsection{X-ray Cavity properties}

In this section, we discuss various properties of the detected X-ray cavities in the SPT-SZ sample. First, we note that the average power, enthalpy and radius of the cavities in the 6 SPT-SZ clusters with convincing X-ray cavities are $\sim5\times10^{45}\ergps$, $\sim6\times10^{59}\erg$ and $\sim16$ kpc, respectively, whereas the average cooling luminosity for these same clusters is $\sim2-3\times10^{45}\ergps$ (depending on which definition we adopt, see Section 4). The X-ray cavities in these 6 SPT-SZ clusters therefore provide on average, enough energy to offset cooling of the hot ICM. This statement remains true even if 1) we include SPT-CLJ0156$-$5541 and SPT-CLJ2342$-$5411 in the calculations and 2), if we only consider the 2 SPT-SZ clusters with ``clear" cavities (SPT-CLJ0509$-$5342 and SPT-CLJ0616$-$5227) where the cavity emission lies $2-3\sigma$ below the surrounding emission. For the latter, the average cavity power, enthalpy and radius would be $\sim0.8\times10^{45}\ergps$, $\sim3\times10^{59}\erg$ and $\sim18$ kpc, respectively, whereas the average cooling luminosity would be $\sim0.1\times10^{45}\ergps$.

We also note that the detection fraction in the sample is $7\%$ since 6 of the 83 SPT-SZ clusters with $Chandra$ observations have visually convincing X-ray cavities. If we only consider the 2 clusters with ``clear" cavities then the fraction drops to $2\%$. This is interesting because local clusters of galaxies with X-ray cavities have typical detection fractions of $20-30\%$ (Dunn $\&$ Fabian 2006; Rafferty et al. 2006; Fabian 2012; Birzan et al. 2012; HL12)\nocite{Dun2006373,Raf2006652,Fab2012,Bir2012427}. However, considering all of the potential selection effects mentioned in Section 6.1, the detection fraction in the SPT-SZ sample is likely a lower limit, and could be consistent with the fraction in local clusters.

\subsection{Evolution of X-ray Cavity properties}

\subsubsection{Implications for AGN feedback}
Interestingly, M13 analysed the cooling properties of the same 83 SPT-SZ clusters with $Chandra$ observations as those studied here, and found that stable long-standing feedback is required to halt cooling of the hot ICM to low temperatures. In agreement with M13, Figs.\ \ref{fig5} and \ref{fig7} show that the enthalpy of X-ray cavities ($pV_{\rm tot}$) does not vary significantly with redshift out to $z\sim0.8$. These results imply that powerful mechanical feedback may have been operating in massive clusters of galaxies for over half of the age of the Universe ($>$7 Gyrs, corresponding to the look-back time since $z\sim0.8$). Newly discovered high-redshift clusters such as WARPJ1415.1+3612 \citep{San2012539} also find that radio-mode feedback must have already been established at $z\sim1$.

In Section 6.3, we found that the average cavity power of the 6 SPT-SZ clusters with convincing cavities was higher than their average cooling luminosity. In fact, they appear to be injecting an excess of $\sim$2$\times10^{45}\ergps$ in heat. According to Fig. \ref{fig6}, this excess is most significant for the clusters in the SPT-SZ sample, in particular, at $z~\gtrsim~0.5$. Assuming that AGN duty cycles remain high and that heating is roughly constant out to $z\sim1$ as suggested by \citet[][]{San2012539}, the excess heat amounts to $\sim1.0$ keV per particle for a total gas mass of 5$\times10^{13}M_{\rm \odot}$. This is similar to the energy needed to explain the excess entropy in clusters \citep{Kai1991383,Voi200577}, and would be $\sim30\%$ larger assuming that the excess heat was being injected out to the onset of cool cores at $z\sim2$ (M13). X-ray cavity powers are also likely only lower limits to the total energy injected by the central AGN, as weak shocks \citep{McN2005433,Fab2006366,For2007665} and sound waves \citep{Fab2003344,San2008390,Bla2011737} contribute to the total energy \citep[see also][]{Nus2006373,Mat2008685}. The excess heat could therefore be even larger. 

Although intriguing, these calculations assume that all of the cavities in the 6 SPT-SZ clusters are real. If we only consider the 2 SPT-SZ clusters with ``clear" cavities (SPT-CLJ0509$-$5342 and SPT-CLJ0616$-$5227), the excess heat would be $\sim$0.7$\times10^{45}\ergps$ (or $0.4$ keV per particle). The estimated excess heat injected per particle also depends on the duty cycle of the energy injection. For local BCGs, this duty cycle has been observed to be high \citep[e.g., $>$60--90\%,][]{Bir2012427,Fab2012}.  However, the lower limit in the SPT-SZ sample is $\sim$11\%, since six of the 52 clusters with signs of cooling have cavities (Fig. \ref{fig3}). Applying this lower limit, we find that the excess in heat is reduced to $0.05-0.1$ keV per particle for cool core clusters. We note that if duty cycles are decreasing with increasing redshift, AGN feedback may no longer be able to suppress all of the ICM cooling. In this case, we would expect to see an average increase in star formation rates for BCGs with increasing redshift. This may explain the unusually high star formation rate in the Phoenix cluster \citep[][]{McD2012Nat,McD2013765}.

In summary, our results imply that the AGN in BCGs may be depositing as little as 0.1 keV per particle or as much as 1.0 keV per particle in excess heat, depending on whether the AGN duty cycles evolve or not between $z=0$ and $z\sim1$.

\subsubsection{Implications for supermassive black hole growth}

Our results show that powerful radio mode feedback may be operating in massive clusters of galaxies for over half of the age of the Universe ($>$7 Gyrs, corresponding to the look-back time since $z\sim0.8$). If we assume once more that the duty cycles of the AGN in BCGs remain high, as they do for local BCGs, then this implies that the supermassive black holes in BCGs may have accreted a substantial amount of mass to power the X-ray cavities. To determine this, we use the following equation relating the jet (or cavity) power ($P_{\rm cav}$) to the black hole accretion rate ($\dot{M}$): 
\begin{equation}
P_{\rm cav} = \eta{\dot{M}}{c^2} \,  ,
\label{eq5}
\end{equation}
where $\eta$ is the efficiency and ${\rm c}$ is the speed of light. We assume that the conversion efficiency between accreted mass and jet power is $\eta=0.1$ \citep{Chu2005363,Mer2008388,Gas2012424}, but stress that if it were lower, then the black holes would need to accrete even more mass. Assuming that the average jet power of SPT-SZ clusters with X-ray cavities is a representative value for massive cool core clusters ($0.8-5\times10^{45}\ergps$), and integrating Eq. \ref{eq5} over 7 Gyrs while assuming nearly constant duty cycles, we find that the supermassive black holes in these BCGs must have accreted $1-6\times10^9M_{\rm \odot}$ in mass to power the radio jets responsible for carving out the observed X-ray cavities. If correct, this would imply that supermassive black hole growth in BCGs may not only be important at earlier times during quasar mode feedback, when the black holes are accreting at rates near the Eddington limit \citep[see][ for a review]{Ale201256}, but also at later times when the black holes are accreting at low rates and driving powerful jetted outflows \cite[see also HL12 and][]{Ma2013763}. We note that if we only consider the lower-limit to the duty cycle ($11\%$ for cool core clusters), then the accreted mass is $0.1-1\times10^9M_{\rm \odot}$ and remains substantial.

\section{Concluding remarks}

We have performed a visual inspection of $Chandra$ X-ray images for 83 SPT-SZ galaxy clusters, finding that 6 contain visually convincing X-ray cavities. These cavities are likely the result of mechanical, or ``radio-mode'', feedback from the central supermassive black hole. This works extends the previous samples of known X-ray cavities to higher redshift (from $z\sim0.5$ to $z\sim0.8$). Interestingly, we find that the Phoenix Cluster ($z=0.596$) is one of these 6 systems with X-ray cavities (see also McDonald et al. in prep), and that its extreme cavity power of $P_{\rm cav}\sim2\times10^{46}\ergps$ rivals those in MS 0735.6+7421 \citep[][]{McN2005433}. On average, the SPT-SZ clusters with detected X-ray cavities have cavity powers of $0.8-5\times10^{45}\ergps$, enthalpies of $3-6\times10^{59}\erg$ and radii of $\sim17$ kpc. We identify two additional systems at $z\sim1$ that contain marginally-detected cavities. We stress the importance of deep $Chandra$ follow-up to confirm and further analyze these structures. The results presented here suggest that powerful mechanical feedback has been operating in massive clusters of galaxies for over half of the age of the Universe ($>$7 Gyrs).

\acknowledgments
JHL is supported by NASA through the Einstein Fellowship Program, grant number PF2-130094, NSERC through the discovery grant and Canada Research Chair programs, as well as FRQNT. Partial support is also provided by the National Science Foundation through grants ANT-0638937 and PLR-1248097, the NSF Physics Frontier Center grant PHY-0114422 to the Kavli Institute of Cosmological Physics at the University of Chicago, the Kavli Foundation, and the Gordon and Betty Moore Foundation. Support for X-ray analysis was provided by NASA through Chandra Award Numbers 12800071, 12800088, and 13800883 issued by the Chandra X-ray Observatory Center, which is operated by the Smithsonian Astrophysical Observatory for and on behalf of NASA. Galaxy cluster research at Harvard is supported by NSF grant AST-1009012 and at SAO in part by NSF grants AST-1009649 and MRI-0723073. This work was supported in part by the U.S. Department of Energy contract numbers DE-AC02-06CH11357 and DE-AC02-76SF00515.

\textit{Facilities}: Blanco (MOSAIC, NEWFIRM), Magellan: Baade (IMACS, FourStar), Magellan: Clay (LDSS3,Megacam), Spitzer (IRAC), South Pole Telescope, Swope (SITe3), Chandra X-ray Observatory (CXO), Molonglo Observatory Synthesis Telescope (MOST).

\bibliographystyle{mn2e}
\bibliography{bibli}
\end{document}

%% file: defn.tex
\newcommand{\apgt}{{\raise-.5ex\hbox{$\buildrel>\over\sim$}}}
\newcommand{\aplt}{{\raise-.5ex\hbox{$\buildrel<\over\sim$}}} 
\newcommand{\Mpc}{\rm\; Mpc}

\newcommand{\km}{\rm\; km}

\newcommand{\s}{\rm\; s}

\newcommand{\keV}{\rm\; keV}

\newcommand{\erg}{\rm\; erg}

\newcommand{\ergps}{\hbox{$\erg\s^{-1}\,$}}

\newcommand{\kmps}{\hbox{$\km\s^{-1}\,$}}

\newcommand{\kmpspMpc}{\hbox{$\kmps\Mpc^{-1}\,$}}

\newcommand{\Omm}{\hbox{$\rm\thinspace \Omega_{m}$}}
\newcommand{\OmL}{\hbox{$\rm\thinspace \Omega_{\Lambda}$}}

